\documentclass[
  aps,
  pra,
  superscriptaddress,
  reprint
]{revtex4-2}

\usepackage{array}[=2016-10-06]
\usepackage[T1]{fontenc}
\usepackage[american]{babel}
\usepackage{newtxtext}

\usepackage{amsmath}
\usepackage{amssymb}
\usepackage{mathtools}
\usepackage{newtxmath}
\usepackage{xfrac}
\usepackage{subdepth}
\usepackage{physics}
\usepackage{mleftright}
\mleftright
\usepackage{bm}

\usepackage{dcolumn}
\usepackage{scalerel}
\usepackage{comment}
\usepackage{graphicx}
\usepackage[HTML]{xcolor}

\usepackage{circledsteps}
\usepackage{tikz}
\newcommand*\crc[1]{%
\hspace{2pt}%
  \begin{tikzpicture}[baseline=(C.base)]
    \node[inner sep=0pt](C) {#1};
    \node[draw=black,solid,circle,inner sep=0pt, minimum size=0.35cm,overlay]at (C.center) {\phantom{#1}};
  \end{tikzpicture}%
  \hspace{2pt}}
\usepackage{booktabs}
\usepackage{tabularx}
\usepackage{etoolbox}
\usepackage[normalem]{ulem}

\usepackage{orcidlink}
\usepackage{hyperref}
\usepackage[all]{hypcap}
\usepackage{adjustbox}
\usepackage{microtype}
\newcolumntype{Y}{>{\raggedright\arraybackslash}X}

\DeclareMathOperator{\sgn}{sgn}

\definecolor{linkblue}{HTML}{2E3092}

\hypersetup{
  colorlinks = true,
  linkcolor  = linkblue,
  citecolor  = linkblue,
  urlcolor   = linkblue,
  filecolor  = linkblue
}

\newcommand{\krondel}{%
    \delta_{\smashoperator[r]{{\textstyle\sum}\limits_{i=1}^{j-1}}%
    {\scriptstyle\overline{n}_i,\ell-1}}%
}
\newcommand{\krondelwide}{%
    \delta_{{\textstyle\sum\nolimits_{i=1}^{j-1}}%
    {\scriptstyle\overline{n}_i,\ell-1}}%
}
\allowdisplaybreaks

\newcommand{\PRLsep}{\begin{center}\resizebox{0.75\linewidth}{1.5pt}{$\bullet$}\end{center}}

\AtBeginDocument{%
  \setlength{\skip\footins}{18pt}%
}

\makeatletter
\patchcmd{\dch@set@one}{H_2}{H_b}
  {}{\errmessage{Subdepth patch 1 failed}}
\patchcmd{\dch@set@one}{H_2}{H_b}
  {}{\errmessage{Subdepth patch 2 failed}}
\patchcmd{\dch@set@one}{+\vrule}{\dagger\vrule}
  {}{\errmessage{Subdepth patch 3 failed}}
\makeatother

\begin{document}
\let\origaddtocontents\addtocontents
\renewcommand{\addtocontents}[2]{%
  \ifstrequal{#1}{toc}{}{\origaddtocontents{#1}{#2}}%
}

\title{Generalized Effective Spin-Chain formalism for multicomponent anyons\texorpdfstring{\\}{ }in one-dimensional optical lattices}

\author{Sagarika Basak\,\orcidlink{0000-0003-2069-644X}}
\email[\!Contact author:~]{basak.sagarika@rice.edu}
\altaffiliation{\href{mailto:basak.sagarika@ou.edu}{basak.sagarika@ou.edu}}
\affiliation{Department of Physics and Astronomy, and the Smalley-Curl Institute, \href{https://ror.org/008zs3103}{Rice University}, Houston, Texas 77251-1892, USA}
\affiliation{%
\adjustbox{max width=\textwidth}{%
Homer L. Dodge Department of Physics and Astronomy,
\href{https://ror.org/02aqsxs83}{The University of Oklahoma},
Norman, Oklahoma 73019, USA%
}}
\affiliation{Center for Quantum Research and Technology, \href{https://ror.org/02aqsxs83}{The University of Oklahoma}, Norman, Oklahoma 73019, USA}

\author{Xi-Wen Guan\,\orcidlink{0000-0001-6293-8529}}
\email[\!Contact author:~]{xiwen.guan@anu.edu.au}
\affiliation{%
\adjustbox{max width=\textwidth}{%
Innovation Academy for Precision Measurement Science and Technology,
\href{https://ror.org/034t30j35}{Chinese Academy of Sciences},
Wuhan 430071, China%
}}
\affiliation{%
\adjustbox{max width=0.99\textwidth}{%
Department of Fundamental and Theoretical Physics, Research School of Physics,
\href{https://ror.org/019wvm592}{Australian National University},
Canberra, ACT 0200, Australia%
}}

\author{Han Pu\,\orcidlink{0000-0002-0018-3076}}
\email[\!Contact author:~]{hpu@rice.edu}
\affiliation{Department of Physics and Astronomy, and the Smalley-Curl Institute, \href{https://ror.org/008zs3103}{Rice University}, Houston, Texas 77251-1892, USA}


\begin{abstract}
We develop a generalized effective spin-chain (GESC) formalism for strongly interacting multicomponent anyons in a one-dimensional (1D) optical lattice. By mapping particle motion onto spinless fermions and spin states onto an ordered spin chain, this framework provides a spin--charge-separated perspective on the physical effects of fractional exchange statistics. \textls[-15]{In the strong-interaction regime, the leading-order charge Hamiltonian becomes independent of the statistical phase; instead, virtual hopping through doubly occupied states directly imprints this phase on the spin-exchange coefficients. The GESC formalism captures ground-state properties in the strong interaction regime of the full Anyon--Hubbard model reported in \href{https://arxiv.org/abs/2609.36554}{S.~Basak, X.-W.~Guan, and H.~Pu, arXiv:2609.36554 (2026)}, and reveals that their statistical dependence is predominantly encoded in the explicit anyonic structure of the observables rather than the underlying spin ground state. This formalism uncovers that, out of equilibrium, successive spin exchanges accumulate direction-dependent phases; the resulting interference drives crossover from dispersive to localized impurity transport and generates inversion-asymmetric propagation at intermediate statistics.} Suppressed expansion persists for identical and distinguishable impurities, with differences in propagation governed by the interplay between impurity identity and interaction anisotropy. Ultimately, GESC offers a versatile, computationally efficient theoretical framework that connects macroscopic dynamics to microscopic virtual exchange processes, conferring a unique spin--charge-separated vantage for resolving how fractional exchange statistics survives strong interactions and manifests distinctly across static and dynamical regimes.
\end{abstract}




\maketitle
\section{\label{sec:introduction}Introduction}
\vspace{-1.5em}Strong interactions fundamentally alter the low-energy physics of one-dimensional (1D) quantum gases. When the interaction energy greatly exceeds the tunneling energy, multiparticle site occupation is energetically suppressed. Then, particle motion governs the charge configuration, while virtual tunneling through the high-energy subspace drives spin exchange. This delineation of energy scales naturally maps a strongly interacting spinor gas onto spinless particles and an effective spin chain. As particles in 1D cannot exchange positions without traversing an energetically costly doubly occupied state, their left-to-right spatial ordering uniquely defines the spin-chain ordering. Related mappings have been widely used to study strongly interacting continuum gases \cite{volosniev_strongly_2014,levinsen_strong-coupling_2015,massignan_magnetism_2015,yang_strongly_2015,yang_bose-fermi_2016,yang_effective_2016}. 

Optical lattices provide a highly controlled platform to realize this strong-interaction regime. Here, confined spinor gases are governed by multicomponent Hubbard models, with superexchange emerging from virtual tunneling \cite{trotzky_time-resolved_2008,greif_short-range_2013,hart_observation_2015,murmann_antiferromagnetic_2015,mazurenko_cold-atom_2017}. Direct simulations become increasingly demanding with system size, component number, and evolution time. Exact diagonalization (ED) is bottlenecked by Hilbert space growth, while density-matrix renormalization-group (DMRG) methods---albeit well suited for 1D systems---struggle with multicomponent gases and long-time dynamics \cite{white_density_1992,schollwock_density-matrix_2011}. Effective frameworks like the $t$-$J$ model simplify the strong-interaction problem by eliminating high-energy configurations, but are largely restricted to spin-$1/2$ fermions or to frozen charge limits \cite{edegger_gutzwillerrvb_2007}.

Fractional exchange statistics adds a profound new dimension to this strong-interaction construction. Anyons continuously interpolate between bosonic$\leftrightarrow$fermionic exchange statistics and can be defined in 1D via generalized exchange relations \cite{haldane_fractional_1991,kundu_exact_1999,batchelor_generalized_2006}. On a lattice, a generalized Jordan--Wigner transformation maps the Anyon--Hubbard model onto a Hubbard model with occupation-dependent tunneling \cite{keilmann_statistically_2011,greschner_anyon_2015,tang_ground-state_2015}. This representation is experimentally relevant because lattice modulation can engineer such tunneling and yield continuously tunable statistical phases \cite{cardarelli_engineering_2016,greschner_probing_2018,kwan_realization_2024}. More recently, many-body anyonic (statistics-dependent) correlations were generated and probed in a strongly interacting 1D gas by spin--charge separation and a mobile impurity \cite{dhar_expcorr_2025,Wang2025Swap}. Single-component anyons already exhibit unconventional ground states and nonequilibrium statistics-dependent behavior \cite{bonkhoff_bosonic_2021,lange_anyonic_2017,arcila-forero_critical_2016,hao_quench_2023}. Multiple components introduce spin degrees of freedom, allowing the statistical phase to impact both charge correlations and spin exchange.

Several theoretical descriptions of 1D anyons are available, including models that are exactly or conditionally solvable by the generalized coordinate Bethe ansatz \cite{kundu_exact_1999,batchelor_generalized_2006,BatchelorGuanOelkers2006,BatchelorGuanKundu2008}. These include the Lieb--Liniger model with fractional exchange statistics \cite{kundu_exact_1999} for contact interactions and the Calogero--Sutherland model \cite{calogero_solution_1971,sutherland_exact_1972,polychronakos_non-relativistic_1989,posske_second_2017} for long-range interactions \cite{auberson_off-diagonal_2000}. Other approaches include a generalization of the Bose--Fermi mapping to an anyon--fermion mapping \cite{girardeau_anyon-fermion_2006} and the Anyon--Hubbard model \cite{bonkhoff_bosonic_2021,keilmann_statistically_2011,greschner_anyon_2015,tang_ground-state_2015,amico_one-dimensional_1998,osterloh_fermionic_2000,osterloh_Schulz-Shastry_2000}. Recent continuum studies generalize Bose--Fermi duality
to interacting boson-type and fermion-type anyons and examine the resulting
off-diagonal correlations and statistics-dependent momentum
distributions
\cite{hidalgosacoto_universal_2025,hidalgosacoto_twoanyons_2025}. Multicomponent (spinor) anyons in an optical lattice have considerable potential for rich physics. However, calculations for multicomponent Anyon--Hubbard models \cite{amico_one-dimensional_1998,osterloh_fermionic_2000,osterloh_Schulz-Shastry_2000} become intractable and nonintuitive at large system size and for large number of internal components, particularly for nonequilibrium dynamics. Occupation-dependent tunneling further couples charge, spin, and statistical degrees of freedom, complicating both computation and physical interpretation. 

To address the analogous strong-interaction problem for conventional spinor gases, we previously introduced a generalized effective spin-chain (GESC) formalism for bosonic and fermionic systems in 1D optical lattices \cite{basak_generalized_2023}. The GESC formalism retains both charge and spin dynamics in coupled spinless-fermion and spin-chain sectors. Related spin--charge representations describe dynamical correlations in lattice gases without double occupancy through impenetrable anyons and an auxiliary spin chain \cite{Gamayun2024}. In contrast, the GESC formalism developed here provides an effective Hamiltonian that retains statistics-dependent virtual spin exchange at large but finite interaction strength. Extending this formalism to fractional exchange statistics thus provides a natural route to an effective description of strongly interacting multicomponent anyons.

Direct multicomponent Anyon--Hubbard calculations reveal an additional physical motivation. In Ref.~\cite{basak_anyon1}, we find a striking dichotomy in the strong-interaction regime: density-based and physical-fermion observables become weakly dependent on the statistical phase, whereas intrinsic anyonic correlations retain robust statistical signatures. The transport dynamics present a parallel distinction: charge-dominated motion can become nearly statistics independent while impurity and momentum-space dynamics remain sensitive. Understanding the origin of this behavior requires delineating the roles of charge motion, spin evolution, and the statistical structure of the observable. The physical origin of these different behaviors is not apparent from the Anyon--Hubbard Hamiltonian alone. In the low-energy subspace, the leading-order charge Hamiltonian contains no statistical phase, yet fractional exchange statistics survives through virtual processes that access doubly occupied states and generate statistics-dependent spin exchange. Statistical dependence can therefore arise through the spin state and/or explicitly through the anyonic operator. A spin--charge-separated representation, such as the GESC formalism, is well suited to distinguish these mechanisms.

Here, we develop the GESC formalism for strongly interacting $N$-component anyons in a 1D optical lattice. We perturbatively decouple the low- and high-energy subspaces of the Anyon--Hubbard model and express the resulting low-energy theory in terms of spinless fermions carrying the charge degrees of freedom and an ordered spin chain carrying the internal components. Fractional exchange statistics enters the spin sector through bond-dependent spin-exchange coefficients $\mathcal{C}_\ell(\theta)$ that encode virtual tunneling through doubly occupied states, while their particle-counting structure maps physical lattice processes onto the corresponding spin-chain bonds. We benchmark the GESC formalism against the two-component Anyon--Hubbard model and then exploit the reduced representation to access larger systems and additional spin components. This provides a natural setting for asking how statistical signatures arise from the many-body state, the structure of the observable, and their interplay, such as those reported in Ref.~\cite{basak_anyon1}. We further extend the formalism to quenched dynamics, using dipole oscillations and the release of single impurities and identical or distinguishable impurity pairs, to probe the roles of charge motion, spin exchange, and impurity composition, and to identify how fractional exchange statistics survives and manifests in the strongly interacting regime.

The remainder of this paper is organized as follows.
Section~\ref{sec:model} introduces the multicomponent Anyon--Hubbard models for fermion- and boson-type anyons and their occupation-dependent tunneling representations.
Section~\ref{sec:effspin} develops the GESC formalism for both classes of anyons and establishes how fractional exchange statistics enters through the bond-dependent spin-exchange coefficients.
Section~\ref{sec:groundstate} examines the ground-state momentum and spin correlations.
Section~\ref{sec:statistical_origin} investigates the origins of their statistical dependence, disentangling their state-mediated and operator-induced sources.
Section~\ref{sec:dynamics} extends the formalism to quenched dynamics, commenting on dipole oscillations and focusing on impurity release.
Finally, Sec.~\ref{sec:summary} provides a summary and outlook. 
Details of the technical derivations, ground-state and dynamical benchmarks, computational-cost comparisons, and additional background information are relegated to Appendices \ref{app:fermionic_derivation}--\ref{app:gesc_efficiency}.

\section{\label{sec:model}System}
\subsection{\label{sec:anyon_model}Multicomponent Anyon--Hubbard model}
\begin{figure}
\centering
\includegraphics{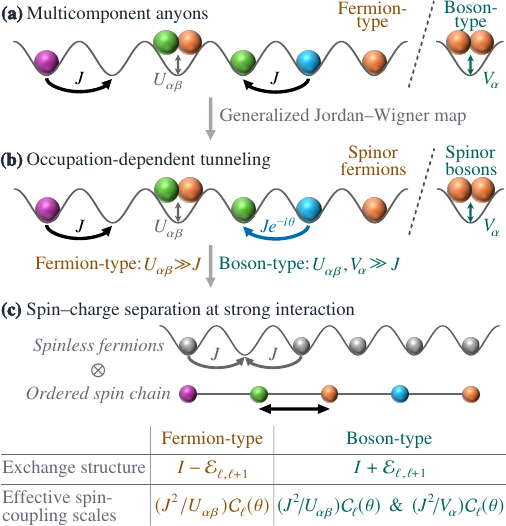}
\caption{\label{fig:schematic}\textbf{Spin--charge separation in strongly interacting anyons.}
(a) Multicomponent anyons with tunneling $J$ and on-site interactions $U_{\alpha\beta}$ and $V_{\alpha}$.
\textls[-10]{Colors denote internal components; split terminal wells} contrast allowed on-site occupations for fermion-/boson-type anyons.
(b) A generalized Jordan--Wigner transformation maps anyons onto multicomponent fermions/bosons with occupation-dependent tunneling.
\textls[-10]{(c) In the strong-interaction regime, the low-energy theory separates} into a spinless-fermion charge sector and an ordered spin chain.}
\end{figure}

We consider a system of $N$-component anyons trapped in a 1D optical lattice \hyperref[fig:schematic]{[Fig.~\ref{fig:schematic}(a)]}. Based on their on-site exchange behavior, we consider two classes of multicomponent anyons: fermion-type anyons, which anticommute on the same site, and boson-type anyons, which commute on the same site. At low temperature in the tight-binding limit, we describe both classes using an $N$-component Anyon--Hubbard model. The fermion-type anyons are governed by the Hamiltonian \cite{osterloh_fermionic_2000,greschner_probing_2018}
\begin{align}
    H_{\mathrm{A:F}} = -J\sum\limits_{i,\alpha}(a_{i,\alpha}^{\dagger}a_{i+1,\alpha}
        + \text{H.c.}) +\sum\limits_{i,\alpha<\beta} U_{\alpha\beta}n_{i,\alpha}n_{i,\beta}\,,
    \label{eqn:FanyonH}
\end{align}
whereas the boson-type anyons are governed by \cite{keilmann_statistically_2011,kwan_realization_2024}
\begin{align}
    \begin{split}
        H_{\mathrm{A:B}} =& -J\sum\limits_{i,\alpha}(a_{i,\alpha}^{\dagger}a_{i+1,\alpha}
            + \text{H.c.}) \\
        & +\sum\limits_{i,\alpha<\beta} U_{\alpha\beta}n_{i,\alpha}n_{i,\beta} +\sum_{i,\alpha}
        \dfrac{V_{\alpha}}{2}n_{i,\alpha} \left(n_{i,\alpha}-1\right)\,.
    \end{split}
    \label{eqn:BanyonH}
\end{align}
In Eqs.~\eqref{eqn:FanyonH} and \eqref{eqn:BanyonH}, $a_{i,\alpha}$ are the anyonic annihilation operators for spin component $\alpha$ at lattice site $i$, $J$ is the tunneling amplitude, and $U_{\alpha\beta}$ is the on-site interaction between anyons of spin components $\alpha$ and $\beta$. Unlike fermion-type anyons, boson-type anyons of the same spin component are allowed to have site occupancy greater than one. Because of this, Eq.~\eqref{eqn:BanyonH} has an additional same-component on-site interaction $V_{\alpha}$. 

The fermion-type anyons follow generalized commutation relations \cite{patu_correlation_2019}
\begin{align}
    \begin{split}
        &a_{j,\alpha}a_{k,\beta}^{\dagger} +e^{-i\theta \sgn(j-k)}
        a_{k,\beta}^{\dagger} a_{j,\alpha} = \delta_{j,k}\delta_{\alpha\beta}\,, \\
        & a_{j,\alpha}a_{k,\beta} + e^{+i\theta \sgn(j-k)} a_{k,\beta}a_{j,\alpha}=0 \,.
    \end{split}
    \label{eqn:FanyonC}
\end{align}
Here, $\theta$ is the statistical phase that anyons acquire upon exchange. For $\theta=0$, the particles act as fermions, whereas for $\theta=\pi$ as pseudobosons. The particles anticommute on the same site and obey generalized commutation relations off site.

The generalized commutation relations for boson-type anyons are given by
\begin{align}
    \begin{split}
        &a_{j,\alpha}a_{k,\beta}^{\dagger} -e^{-i\theta \sgn(j-k)}
        a_{k,\beta}^{\dagger} a_{j,\alpha} = \delta_{j,k}\delta_{\alpha\beta}\,, \\
        & a_{j,\alpha}a_{k,\beta} - e^{+i\theta \sgn(j-k)} a_{k,\beta}a_{j,\alpha}=0 \,.
    \end{split}
    \label{eqn:BanyonC}
\end{align}
In this case, for $\theta=0$, the particles act as bosons, whereas for $\theta=\pi$ as pseudofermions. The particles therefore commute on the same site and obey generalized commutation relations off site.
We take the statistical phase to be spin independent and identical for all components. The statistical phase can be generalized such that exchanges between different components acquire different phases, but that will be a future consideration.

\subsection{\label{sec:anyon_fermion_mapping}Anyon--fermion mapping}
We map the multicomponent fermion-type anyons onto multicomponent fermions via the generalized Jordan--Wigner transformation \cite{greschner_probing_2018}
\begin{align}
    a_{i,\alpha} = \exp\left(
        -i\theta \textstyle\sum\limits_{l<i}n_l
    \right)c_{i,\alpha}\,,
    \label{eqn:FJWMap}
\end{align}
where $c_{i,\alpha}$ annihilates a local fermion of component $\alpha$ at site $i$, and $n_l=\sum_\beta n_{l,\beta}$ is the total occupation of site $l$. The Jordan--Wigner string makes the representation of the anyonic operator nonlocal and ensures the generalized commutation relations in Eq.~\eqref{eqn:FanyonC}; see Appendix~A in Ref.~\cite{basak_anyon1}.

The mapping in Eq.~\eqref{eqn:FJWMap} transforms Eq.~\eqref{eqn:FanyonH} into the multicomponent Fermi--Hubbard Hamiltonian governing the parent physical fermions $c_{i,\alpha}$:
\begin{align}
    H_{\mathrm{F}} = &-J\sum\limits_{i,\alpha} \left(
        \begin{aligned}
        &c_{i,\alpha}^{\dagger}e^{-i\theta n_{i}}c_{i+1,\alpha}\\
        &+ c_{i+1,\alpha}^{\dagger}e^{+i\theta n_{i}} c_{i,\alpha}    
        \end{aligned}
    \right)
    +\smashoperator{\sum\limits_{i,\alpha<\beta}} U_{\alpha\beta}
    n_{i,\alpha}n_{i,\beta}\,.
\label{eqn:fermiH}
\end{align}
This generalized Fermi--Hubbard Hamiltonian demonstrates occupation-dependent tunneling $J\exp(\pm i\theta n_{i})$ and on-site interaction $U_{\alpha\beta}$ as shown in \hyperref[fig:schematic]{Fig.~\ref{fig:schematic}(b)}. 

\subsection{\label{sec:anyon_boson_mapping}Anyon--boson mapping}
We map the multicomponent boson-type anyons onto multicomponent bosons via the generalized Jordan--Wigner transformation \cite{keilmann_statistically_2011}
\begin{align}
    a_{i,\alpha} = \exp\left(
        -i\theta \textstyle\sum\limits_{l<i}n_l
    \right)b_{i,\alpha}\,,
    \label{eqn:BJWMap}
\end{align}
where $b_{i,\alpha}$ annihilates a local boson of component $\alpha$ at site $i$. As was the case for fermion-type anyons, the Jordan--Wigner string makes the representation of the anyonic operator nonlocal, enforcing the generalized commutation relations in Eq.~\eqref{eqn:BanyonC}.

Applying the mapping in Eq.~\eqref{eqn:BJWMap} to the Anyon--Hubbard Hamiltonian in Eq.~\eqref{eqn:BanyonH} yields the generalized multicomponent Bose--Hubbard Hamiltonian governing the parent physical bosons $b_{i,\alpha}$:
\begin{align}
    \begin{split}
        H_{\mathrm{B}} = &-J\sum\limits_{i,\alpha} \left(
            b_{i,\alpha}^{\dagger}e^{-i\theta n_{i}}b_{i+1,\alpha}
            + b_{i+1,\alpha}^{\dagger}e^{+i\theta n_{i}} b_{i,\alpha}
        \right) \\
        &+\smashoperator{\sum\limits_{i,\alpha<\beta}} U_{\alpha\beta}
        n_{i,\alpha}n_{i,\beta}
        +\smashoperator{\sum\limits_{i,\alpha}} \dfrac{V_{\alpha}}{2}
        n_{i,\alpha}\left(n_{i,\alpha}-1\right)\,.
    \end{split}
    \label{eqn:boseH}
\end{align}
In addition to the occupation-dependent tunneling $J\exp(\pm i\theta n_{i})$ and on-site intercomponent interaction $U_{\alpha\beta}$ present in Eq.~\eqref{eqn:fermiH}, the generalized Bose--Hubbard Hamiltonian contains on-site intracomponent interaction $V_{\alpha}$ \hyperref[fig:schematic]{[Fig.~\ref{fig:schematic}(b)]}.

For both mappings, the tunneling amplitudes in Eqs.~\eqref{eqn:fermiH} and \eqref{eqn:boseH} acquire a density-dependent Peierls-type phase that breaks reflection parity. The generalized Jordan--Wigner string thus transforms nearest-neighbor anyonic hopping into correlated nearest-neighbor fermionic/bosonic hopping. As the density operators are unchanged under this transformation, the interaction terms retain the same form as in the corresponding Anyon--Hubbard Hamiltonians. Despite the nonlocal character of the mapping, the resulting mapped Hamiltonians contain only local operators. Multicolor lattice-depth modulation in a tilted lattice provides a route to engineer the required occupation-dependent tunneling~\cite{cardarelli_engineering_2016,kwan_realization_2024}.

\section{\label{sec:effspin}Generalized Effective Spin-Chain (GESC) Hamiltonian}
In Ref.~\cite{basak_anyon1}, we study the ground-state and dynamical properties of fermion-type anyons using the Anyon--Hubbard Hamiltonian. In the strong-interaction regime ($U_{\alpha\beta}, V_{\alpha} \gg J$), spinor anyons can be studied using the GESC formalism, which maps the spinor gas onto spinless fermions and a spin chain \hyperref[fig:schematic]{[Fig.~\ref{fig:schematic}(c)]} governed by an effective GESC Hamiltonian. Beyond a simpler and highly efficient tool to study spinor anyons, the spin--charge-separated structure of the GESC formalism confers a unique vantage into the origin of the fractional-statistics-dependent behavior in the strong-interaction regime.

\subsection{\label{sec:spin_charge_mapping}Spin--charge mapping}
We generalize the mapping previously developed for 1D spinor bosonic/fermionic gases \cite{basak_generalized_2023} to include fractional exchange statistics, so as to map the spinor anyons governed by Eqs.~\eqref{eqn:FanyonH} and \eqref{eqn:BanyonH} onto spinless fermions and a spin chain. This mapping exploits a property of 1D systems: the configuration space of $M$ particles can be decomposed into $M!$ spatial sectors, and the total wave function can be reconstructed from the wave function in one sector. For spinor bosonic/fermionic gases, the total wave function defined using the wave function of spatial sector $1$ ($\Psi^1$; $x_1<x_2<\ldots<x_M$) \cite{yang_bose-fermi_2016} is
\begin{align}
    \Psi\left(
        \begin{aligned}
            x_1,\,x_2,\,\ldots,\,x_M,\\
            \sigma_1,\,\sigma_2,\,\ldots,\,\sigma_M
        \end{aligned}
    \right) = \sum\limits_P (\pm 1)^P P\bm{\Bigg(}\Psi^1\left(
        \begin{aligned}
            x_1,\,x_2,\,\ldots,\,x_M,\\
            \sigma_1,\,\sigma_2,\,\ldots,\,\sigma_M
        \end{aligned}
    \right)\bm{\Bigg)}\,,
\label{eqn:Otwfn}
\end{align}
where $x_k$ and $\sigma_k$ denote the position and spin of the $k^{\rm th}$ particle, respectively, and $P$ is a permutation. In spatial sector $1$, the wave function is written as a product of charge ($\varphi$) and spin ($\chi$) wave functions, with superposition coefficients $A_{a,b}$ \cite{yang_bose-fermi_2016}:
\begin{align}
    \Psi^1\left(
        \begin{aligned}
            x_1,\,x_2,\,\ldots,\,x_M,\\
            \sigma_1,\,\sigma_2,\,\ldots,\,\sigma_M
        \end{aligned}
    \right) = \smashoperator{\sum\limits_{a,b}} A_{a,b} \left[
        \begin{aligned}
            &\varphi_a(x_1,\,x_2,\,\ldots,\,x_M)\\
            &\times\chi_b(\sigma_1,\,\sigma_2,\,\ldots,\,\sigma_M)
        \end{aligned}
    \right]\,.
\label{eqn:Oswfn}
\end{align}

For both fermion- and boson-type anyons, the wave function must satisfy the generalized exchange symmetry imposed by the corresponding anyonic commutation relations [Eqs.~\eqref{eqn:FanyonC} or \eqref{eqn:BanyonC}, respectively] \cite{hao_ground-state_2009}
\begin{align}
    \begin{split}
        &\Psi_{\mathrm{A}}\left(
            \begin{aligned}
                &x_1,\,\ldots,\,x_j,\,\ldots,\,x_l,\,\ldots,\,x_M,\\
                &\sigma_1,\,\ldots,\sigma_j,\,\ldots,\sigma_l,\,\ldots,\sigma_M
            \end{aligned}
        \right) \\
        &{}= \pm e^{-i\kappa} \Psi_{\mathrm{A}}\left(
            \begin{aligned}
                &x_1,\,\ldots,\,x_l,\,\ldots,\,x_j,\,\ldots,\,x_M,\\
                &\sigma_1,\,\ldots,\sigma_l,\,\ldots,\sigma_j,\,\ldots,\sigma_M
            \end{aligned}
        \right)\,,
    \end{split}
    \label{eqn:Wavesym}
\end{align}
where the plus sign corresponds to boson-type anyons and the minus sign corresponds to fermion-type anyons. The phase acquired upon exchange of anyons is
\begin{align}
    \kappa = \theta\left(
        \smashoperator[r]{\sum\limits_{k=j+1}}^l \sgn(x_j-x_k)
        - \smashoperator{\sum\limits_{k=j+1}^{l-1}} \sgn(x_l-x_k)
    \right) \,.
\end{align}\smallskip

In spatial sector 1, we express the anyonic wave function satisfying the generalized exchange symmetry condition in terms of the reference wave function [defined in Eq.~\eqref{eqn:Oswfn}] as
\begin{align}
    \begin{split}
        &(\Psi_{\mathrm{A}})^1\left(
            \begin{aligned}
                x_1,\,x_2,\,\ldots,\,x_M,\\
                \sigma_1,\,\sigma_2,\,\ldots,\,\sigma_M
            \end{aligned}
        \right) \\
        &{}= \exp(-i\dfrac{\theta}{2}\smashoperator[r]{\sum\limits_{\mathrlap{\!\!\!\!\!1\leq j<k \leq M}}}
            \sgn(x_j-x_k))\Psi^1\left(
            \begin{aligned}
                x_1,\,x_2,\,\ldots,\,x_M,\\
                \sigma_1,\,\sigma_2,\,\ldots,\,\sigma_M
            \end{aligned}
        \right)\,.
    \end{split}
    \label{eqn:twfn}
\end{align}
The wave function in any other spatial sector can be obtained by applying the corresponding permutation operator and statistical exchange factor to the wave function
\begin{align}
    \begin{split}
        (\Psi_{\mathrm{A}})^P& = P\bm{(}(\Psi_{\mathrm{A}})^1\bm{)} \\
        &= \left(\pm 1\right)^{P} \exp(-i\dfrac{\theta}{2}
            \smashoperator[r]{\sum\limits_{\mathrlap{\!\!\!\!\!1\leq j<k \leq M}}} \sgn(x_{P(j)}-x_{P(k)}))P
        \left(\Psi^1\right)\,.
    \end{split}
\end{align}
Therefore, the total wave function of spinor anyons is
\begin{widetext}
\vspace{-9pt}
\begin{align}
    \begin{split}
        \Psi_{\mathrm{A}}\left(
            \begin{aligned}
                &x_1,\,x_2,\,\ldots,\,x_M,\\
                &\sigma_1,\,\sigma_2,\,\ldots,\,\sigma_M
            \end{aligned}
        \right) = \sum\limits_P (\pm 1)^P \exp(-i\dfrac{\theta}{2}
            \smashoperator[r]{\sum\limits_{\mathrlap{\!\!\!\!\!1\leq j<k \leq M}}}\sgn(x_{P(j)}-x_{P(k)}))P
        \bm{\Bigg(}\Psi^1\left(
            \begin{aligned}
                &x_1,\,x_2,\,\ldots,\,x_M,\\
                &\sigma_1,\sigma_2,\ldots,\sigma_M
            \end{aligned}
        \right)\bm{\Bigg)}\,.
    \end{split}
    \label{eqn:Xtwfn}
\end{align}
\end{widetext}
Equations~\eqref{eqn:Xtwfn} and~\eqref{eqn:twfn} show that, in 1D, the anyonic wave function is fully determined by its wave function in one spatial sector together with the statistical phase. Within that sector, the wave function can be expressed as a direct product of the charge ($\varphi$) and spin ($\chi$) wave functions. The spin--charge mapping in the continuum depends on this construction. For the strongly interacting lattice gas considered here, however, we need not construct the full anyonic wave function from all spatial sectors. The spin--charge separation of the wave function instead emerges naturally from the strong-interaction expansion: the projected tunneling governs the charge sector and determines $\varphi$, whereas virtual excitations through the high-energy manifold generate the effective spin exchange and determine $\chi$. Thus, the continuum wave function motivates the mapping, while the lattice formulation realizes the spin--charge separation directly through perturbation theory.

\subsection{\label{sec:fermionic_gesc}Fermion-type anyons: GESC Hamiltonian}
We utilize the three-step protocol previously developed for strongly interacting spinor bosons/fermions \cite{basak_generalized_2023}: 
(1) decouple the low- and high-energy subspaces of the Hilbert space to obtain an effective Hamiltonian through a perturbative expansion up to second order, 
(2) reformulate the effective Hamiltonian in a spin--charge-separated form, and 
(3) apply degenerate perturbation theory. 
We thereby derive the GESC Hamiltonian for $N$-component fermion-type anyons acting on the mapped system of spinless fermions and a spin chain (see Appendix~\ref{app:fermionic_derivation}):
\begin{align}
    \begin{split}
        H_{\mathrm{SC:F}}=E_0 -J^2\sum\limits_{\ell=1}^{M-1}\mathcal{C}_\ell(\theta)\left(
            \sum\limits_{\alpha<\beta}^{N}\dfrac{(I-\mathcal{E}_{\ell,\ell+1})}{U_{\alpha\beta}}
            \hat{P}_{\ell,\ell+1}^{\alpha\beta}
        \right)\,.
    \end{split}
    \label{eqn:HSCF}
\end{align}
Here, $E_0$ is the ground-state energy of spinless fermions on the same lattice, while $\ell$ labels the $\ell^\text{th}$ particle in the fixed left-to-right ordering and hence a site of the effective spin chain, rather than a physical lattice site. $M$ is the number of anyons and therefore the number of spins in the chain. The spin-exchange operator $\mathcal{E}_{\ell,\ell+1}$ swaps neighboring spins, while $\hat{P}_{\ell,\ell+1}^{\alpha\beta}$ projects them onto the $(\alpha,\beta)$ component subspace.

Equation~\eqref{eqn:HSCF} describes bond-dependent superexchange between neighboring spins. Within the projected $(\alpha,\beta)$ subspace, for each neighboring pair, $(I-\mathcal{E}_{\ell,\ell+1})$ represents the spin exchange generated by virtual double occupancy, with energy scale $J^2/U_{\alpha\beta}$. The charge configuration and fractional exchange statistics fully enter through the bond-dependent coefficient
{\medmuskip=0mu
 \thinmuskip=0mu
\begin{align}
    \begin{split}
        \mathcal{C}_\ell(\theta) &= \expval{
            \sum\limits_{j=1}^{L} \krondel \left(
                \text{\small$
                    \begin{aligned}
                        2\bar{n}_j \bar{n}_{j+1} &- e^{+i\theta}
                        f_{j+2}^{\dagger} \bar{n}_{j+1} f_j \\
                        &- e^{-i\theta}f_{j-1}^{\dagger} \bar{n}_j
                        f_{j+1}
                    \end{aligned}
                    $}
            \right)
        }{\varphi}\,,
    \end{split}
    \label{eqn:HSCF1}
\end{align}
}where $f_{j}$ and $\bar{n}_{j}$ are the spinless fermionic annihilation and number operators, respectively, $j$ labels a physical lattice site, and $L$ is the number of lattice sites. The Kronecker delta acts as a left-to-right particle counter: the condition $\sum_{i=1}^{j-1}\bar{n}_i=\ell-1$ identifies the spinless fermion associated with physical site $j$ as the $\ell^\text{th}$ spin of the spin chain. It thus provides the correspondence between the physical lattice and the spin chain.

The three contributions to $\mathcal{C}_\ell(\theta)$ have distinct physical interpretations. The term $2\bar{n}_j\bar{n}_{j+1}$ describes conventional nearest-neighbor superexchange when the two ordered particles occupy adjacent physical sites. The terms proportional to $e^{\pm i\theta}$ describe opposite directions of occupied-neighbor-assisted three-site tunneling through an intermediate doubly occupied state. Consequently, although the leading-order charge Hamiltonian is that of spinless fermions and has no statistical phase, the fractional exchange statistics modifies the effective spin dynamics through the bond-dependent $\mathcal{C}_\ell(\theta)$. For a selected component pair $(\alpha,\beta)$, $(I-\mathcal{E}_{\ell,\ell+1})$ vanishes on a symmetric neighboring-spin state and is nonzero on its antisymmetric component. Thus, for repulsive interactions and positive $\mathcal{C}_\ell(\theta)$, the effective exchange lowers the energy of antisymmetric spin pairs and favors antiferromagnetic ordering.\medskip

\subsection{\label{sec:bosonic_gesc}Boson-type anyons: GESC Hamiltonian}
Utilizing the same three-step protocol for boson-type anyons yields the GESC Hamiltonian acting on the mapped system of spinless fermions and a spin chain (see Appendix~\ref{app:bosonic_derivation}) as
\begin{widetext}
\begin{align}
    \begin{split}
        H_{\mathrm{SC:B}} = E_0- J^2\sum\limits_{\ell}\mathcal{C}_{\ell}(\theta)\left\{
            \sum\limits_{\alpha<\beta}\dfrac{1}{U_{\alpha\beta}}\left[
                \begin{aligned}
                    \left(I+\mathcal{E}_{\ell,\ell+1}\right)
                    &+ \left(\dfrac{\mu_{\alpha;\beta}+\mu_{\beta;\alpha}}{2}-1\right) \left(
                        I+S_{\ell,\left(\alpha\beta\right)}^zS_{\ell+1,\left(\alpha\beta\right)}^z
                    \right)\\
                    &+ \dfrac{\mu_{\beta;\alpha}-\mu_{\alpha;\beta}}{2}\left(
                        S_{\ell,\left(\alpha\beta\right)}^z+S_{\ell+1,\left(\alpha\beta\right)}^z
                    \right)
                \end{aligned}
            \right] \hat{P}^{\alpha\beta}_{\ell,\ell+1}
        \right\}\;,
    \end{split}
    \label{eqn:HSCB}
\end{align}
\end{widetext}
where, in addition to the quantities defined for fermions, $\mu_{\alpha;\beta} = U_{\alpha,\beta}/[(N-1)V_{\alpha}]$, and $S_{\ell,\left(\alpha\beta\right)}^z = n_{\ell,\beta}-n_{\ell,\alpha}$ is the Pauli operator along $z$ in the two-component $(\alpha,\beta)$ subspace. The bond-dependent spin-exchange coefficients $\mathcal{C}_\ell(\theta)$ are the same as those defined for fermion-type anyons in Eq.~\eqref{eqn:HSCF1}. 

The combination of $E_0$ and the contribution  $\propto(I+\mathcal{E}_{\ell,\ell+1})$ in Eq.~\eqref{eqn:HSCB} has the same structure as the fermion-type GESC Hamiltonian [Eq.~\eqref{eqn:HSCF}], except that $(I-\mathcal{E}_{\ell,\ell+1})$ is replaced by $(I+\mathcal{E}_{\ell,\ell+1})$. For a selected component pair $(\alpha,\beta)$, now that it is instead an identity \textit{plus} a spin exchange operation, it selects the symmetric nearest-neighbor spin states, reflecting the on-site exchange symmetry of boson-type anyons. Consequently, for repulsive interactions and positive $\mathcal{C}_\ell(\theta)$, this operation lowers the energy of symmetric neighboring-spin configurations and favors ferromagnetic ordering.

The remaining two terms inside the square brackets in Eq.~\eqref{eqn:HSCB} arise from the intracomponent interactions $V_{\alpha}$, which have no counterpart for fermion-type anyons. The first of these terms, $\propto (I+S_{\ell,\left(\alpha\beta\right)}^zS_{\ell+1,\left(\alpha\beta\right)}^z)$, describes an additional ordering between neighboring spins along $z$ in the two-component $(\alpha,\beta)$ subspace. Its strength is determined by the relative magnitudes of the inter- and intracomponent interactions through the term $\abs{(\mu_{\alpha;\beta}+\mu_{\beta;\alpha})/2-1}$. This term therefore captures the interaction anisotropy and can change the preferred magnetic ordering. For two components with spin-independent intracomponent interaction, its coefficient $\propto (U_{\alpha\beta}/V-1)$ favors ferromagnetic ordering for $U_{\alpha\beta}>V$, antiferromagnetic ordering for $U_{\alpha\beta}<V$, and vanishes at $U_{\alpha\beta}=V$. 

The final term is nonzero only when the intracomponent interactions are spin-component dependent, $V_{\alpha}\neq V_{\beta}$. It acts as an effective longitudinal field in the $(\alpha,\beta)$ subspace and energetically favors one component over the other. Its strength and orientation are determined by the difference between $\mu_{\alpha;\beta}$ and $\mu_{\beta;\alpha}$, and hence by the difference between the corresponding intracomponent interaction strengths.

\subsection{\label{sec:prelude}Prelude to the physics}
In the strongly interacting regime, the leading-order charge Hamiltonian is that of spinless fermions and is independent of the statistical phase. Fractional exchange statistics instead enters the effective spin dynamics through the bond-dependent spin-exchange coefficients $\mathcal{C}_\ell(\theta)$, originating from the occupied-neighbor-assisted three-site tunneling. This tunneling process contributes to the effective spin-chain couplings that govern the state-mediated statistical dependence. Intrinsic anyonic observables additionally retain explicit statistical strings, as discussed in Sec.~\ref{sec:statistical_origin}.

The GESC Hamiltonian developed here has a similar form to that for fermions/bosons \cite{basak_generalized_2023}, with an additional $\theta$ dependence emerging in the spin-exchange coefficients. Taking $\theta=0$, it recovers the results of Ref.~\cite{basak_generalized_2023}. With the separation of spin and charge degrees of freedom, the GESC formalism provides a powerful tool to investigate strongly interacting spinor gases, permitting consideration of larger system sizes, higher numbers of spin components, and longer time evolution. Additionally, the spin--charge-separated form enables the analysis of the origin of fractional-statistics-dependent behavior. 

These observations provide the physical basis for the analyses that follow. We first examine the ground-state properties to identify how fractional exchange statistics manifests in charge- and spin-sensitive observables and to establish the validity of the GESC formalism, before turning to the nonequilibrium dynamics of strongly interacting spinor anyons.

\section{\label{sec:groundstate}Ground-state properties}
The intrinsic momentum distribution and spin structure factor provide complementary probes of one-body density and spin correlations, respectively, and reveal characteristic signatures of fractional exchange statistics in 1D anyonic systems. Building on our Anyon--Hubbard study of $N$-component anyons \cite{basak_anyon1}, we use the GESC formalism to investigate the ground-state statistical signatures in the strongly interacting regime and distinguish operator-induced statistical dependence, arising from the explicit anyonic structure of the observable, from state-mediated statistical dependence, entering through the many-body state. Specifically, we study the charge and spin correlations and their associated momentum distribution and spin structure factor, respectively, using density matrix renormalization group (DMRG) with open boundary conditions. We first define the relevant operators in the spin--charge-separated representation.
\paragraph*{Momentum Distribution}\!\!\!is defined as the occupation operator of the anyons in the momentum space and expressed, assuming lattice spacing $1$, as \cite{manmana_su_2011}
\begin{align}
    n_k = \dfrac{1}{N}\sum\limits_{\sigma} \expval{a_{k,\sigma}^{\dagger}a_{k,\sigma}}
    = \dfrac{1}{L}\sum\limits_{m,n=1}^{L} C(m,n) e^{ik(m-n)}\,.
\end{align}
It is the Fourier component of the one-body density correlation:
\begin{align}
    C(m,n) = \dfrac{1}{N}\sum\limits_{\sigma}C_{\sigma}(m,n) = \dfrac{1}{N}\sum\limits_{\sigma}
    \expval{a_{m,\sigma}^{\dagger}a_{n,\sigma}}\,.
\end{align}
We have detailed the spin--charge-separated implementation of this operator for a lattice system in Ref.~\cite{basak_generalized_2023}. In the mapped representation here, we write it as
\begin{align}
    &C^{n>m}(m,n) \\
    &= \smashoperator[r]{\sum_{m_p,n_p=1}^{M}} (\pm 1)^{\zeta}e^{-i\theta \zeta}
    \dfrac{D^{m,n}_{m_p,n_p}}{N} \expval{\mathcal{E}_{m_p,m_p+1}\ldots\mathcal{E}_{n_p-1,n_p}}{\chi} \notag\\
    &C^{n<m}(m,n) \\
    &= \smashoperator[r]{\sum_{m_p,n_p=1}^{M}} (\pm 1)^{-\zeta}e^{+i\theta \zeta}
    \dfrac{D^{m,n}_{m_p,n_p}}{N} \expval{\mathcal{E}_{m_p,m_p-1}\ldots\mathcal{E}_{n_p+1,n_p}}{\chi}\,,\notag
\end{align}
where the plus sign is for boson-type and the minus sign is for fermion-type anyons, and the exponent $\zeta = \abs{n_p-m_p}$. The relevant charge degrees of freedom are contained in the factor
\begin{align*}
    D^{m,n}_{m_p,n_p} &= \expval{
        \delta_{\sum_{i_0=1}^{m-1}n_{i_0}}^{m_p-1}
        \delta_{\sum_{i_1=1}^{n-1}n_{i_1}}^{n_p-1}
        f_{m}^{\dagger}f_{n}
    }{\varphi}\,,\\
    \text{where } \;\delta_{\sum_{i_0=1}^{m-1}n_{i_0}}^{m_p-1} &=
    \begin{cases}
        1,& \text{if } \sum_{i_0=1}^{m-1}n_{i_0} = m_p-1\\
        0,& \text{otherwise}\,.
    \end{cases}
\end{align*}
The $\delta$ operator ensures that the $n^{\text{th}}$ site is occupied by the $n_{p}^{\text{th}}$ spin, and after the hop the $m^{\text{th}}$ site is occupied by the $m_{p}^{\text{th}}$ spin.

\paragraph*{Structure Factor}\!\!\!associated with the two-body spin correlations $S(m,n)$ is expressed as
\begin{align}
    \begin{split}
        S(k) = \dfrac{1}{L}\sum\limits_{m,n=1}^{L} S(m,n) e^{ik(m-n)}\,,
    \end{split}
    \label{eqn:diaSSF}
\end{align}
Focusing on the diagonal spin correlations, the two-body spin correlation is defined as
{\medmuskip=0mu
 \thinmuskip=0mu
\begin{align}
    \begin{split}
        S(m,n) = \dfrac{1}{N(N-1)} \sum\limits_{\alpha \neq\beta} \left(
            \expval{n_{m,\alpha}n_{n,\alpha}}-\expval{n_{m,\alpha}n_{n,\beta}}
        \right)\,.
    \end{split}
    \label{eqn:tbdsc}
\end{align}
}In the mapped system, we express the spin correlations entering the structure factor as
\begin{align}
    \begin{split}
        &S(m,n) = \smashoperator[l]{\sum\limits_{m_p,n_p=1}^{M}} \dfrac{D^{m,n}_{s,m_p,n_p}}{N(N-1)}
        \sum\limits_{\alpha\neq\beta} \left(
            \text{\small$
                \begin{aligned}
                    &\expval{n_{m_p,\alpha}n_{n_p,\alpha}}{\chi} \\
                    &-\expval{n_{m_p,\alpha}n_{n_p,\beta}}{\chi}
                \end{aligned}
                $}
        \right)\,, \\
        &\text{where~} D^{m,n}_{s,m_p,n_p} = \expval{
            \delta_{\sum_{i_0=1}^{m-1}n_{i_0}}^{m_p-1} \delta_{\sum_{i_1=1}^{n-1}n_{i_1}}^{n_p-1} n_m n_n
        }{\varphi}\,.
    \end{split}
\end{align}

\subsection{\label{sec:multicomponent_groundstate}Multicomponent ground states}
Before turning to larger multicomponent systems, we benchmark the GESC formalism against full Anyon--Hubbard calculations in Appendix~\ref{app:gesc_realspace_benchmarks}. The comparison establishes that the GESC formalism reproduces the relevant strong-interaction charge and spin correlations. Having established this benchmark, we use the GESC formalism to explore larger systems and additional spin components beyond those readily accessible in the full Anyon--Hubbard treatment.

We examine the momentum distribution and spin structure factor in GESC at $L=60$, taking component-independent intercomponent interactions $U_{\alpha\beta}=U=20J$ and, for boson-type anyons, intracomponent interactions $V_\alpha = V \in \{10,\;  40\}J$. Throughout this section, we restrict the ground state to the spin-balanced sector, with equal populations $M_\sigma=M/N$ in each component. The statistical phase is sampled at 21 values, $\theta/\pi \in \{0,\;0.05,\;\ldots,\;1\}$. The momentum distributions are compared at fixed population per component, $M/N=M_\sigma=6$, while the total particle number $M=6N$ increases with $N$. The spin structure factors are compared at fixed total population, $M=36$, such that the population per component $M_\sigma=36/N$ decreases with increasing $N$. These two parameter sequences probe complementary aspects of the $N$ dependence. The lattice spacing is set to unity.

\subsubsection{\label{sec:gesc_main_momentum}Momentum distribution}

\begin{figure}
\centering
\includegraphics{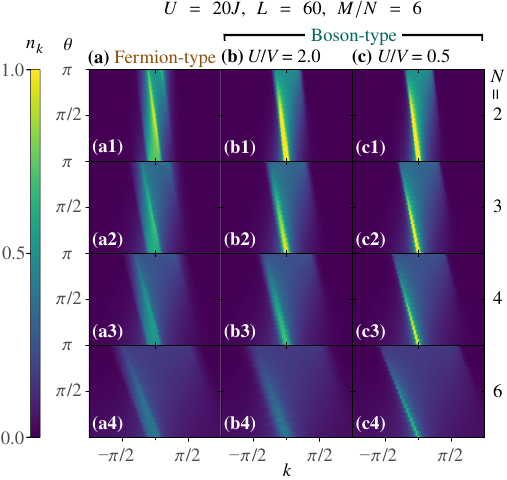}
\caption{\label{fig:gesc_main_nk_heatmaps}\textbf{Statistical-phase-driven asymmetric reconstruction of intrinsic momentum distributions.} Intrinsic anyonic momentum distribution $n_k(\theta)$, averaged over $N$ internal components, from GESC. The statistical phase, $\theta$, shifts the dominant features and redistributes spectral weight. At fixed population per component, increasing $N$ also increases the filling. Colors saturate for $n_k \geq 1$.}
\end{figure}

\hyperref[fig:gesc_main_nk_heatmaps]{Figure~\ref{fig:gesc_main_nk_heatmaps}} presents the component-averaged intrinsic anyonic momentum distribution $n_k(\theta)$ for $N \in \{2,\;3,\;4,\;6\}$. For fermion-type anyons \hyperref[fig:gesc_main_nk_heatmaps]{[Fig.~\ref{fig:gesc_main_nk_heatmaps}(a)]}, intermediate $\theta$ produces unequal finite-momentum features, whose positions and sharpness reflect the oscillation wavevectors and decay exponents of the one-body density correlations \cite{santos_quantum_2012,patu_correlation_2019}. The corresponding Anyon--Hubbard analysis is developed in Appendices~C and E of Ref.~\cite{basak_anyon1}. Reflection symmetry is recovered at the pseudobosonic endpoint, although the resulting profile differs from the original fermionic distribution.

For boson-type anyons \hyperref[fig:gesc_main_nk_heatmaps]{[Fig.~\ref{fig:gesc_main_nk_heatmaps}(b,c)]}, increasing $\theta$ transforms the central peak into unequal finite-momentum features and eventually into a symmetric, shell-like distribution at the pseudofermionic endpoint \cite{hao_ground-state_2008}. This progression occurs for all $N$ and both interaction ratios, with the peak remaining more pronounced for $U/V=0.5$ at larger $N$. Within the spin-balanced sector, at $U/V=2$, the favored ferromagnetic ordering manifests as same-spin domains, whereas $U/V=0.5$ favors antiferromagnetic ordering. This allows the latter to retain larger overlaps under the spin permutations entering $C(m,n)$, and thus result in a pronounced peak.   

Across the rows, increasing $N$ at fixed $M/N$ increases filling while strong interactions exclude multiple occupancy, consistent with the reduced peak weight and broader momentum distribution. For spin-balanced boson-type anyons at $U/V=2$, domain formation could provide an additional broadening mechanism. If each component forms a domain, increasing $N$ adds more domains within the same lattice. Their smaller spatial extent could restrict one-body density correlations to shorter distances, further broadening the momentum peaks.

To characterize the reconstruction of the momentum distributions induced by fractional exchange statistics, as seen in \hyperref[fig:gesc_main_nk_heatmaps]{Fig.~\ref{fig:gesc_main_nk_heatmaps}}, we use four complementary diagnostics introduced in Appendix~B of
Ref.~\cite{basak_anyon1}. The normalized momentum distribution $p_{\theta}(k) = {n_k(\theta)}/{\textstyle\int n_k(\theta)\,dk}$,
such that $\int p_{\theta}(k) dk =1$, is used to obtain the redistribution $R$, mirror asymmetry $\mathcal{A}_{\mathrm{M}}$, mean momentum $\bar{k}$, and momentum width $\sigma_k$, given by
\begin{align}
  R(\theta)
  &=
  \frac{1}{2}\textstyle\int \abs{p_{\theta}(k)-p_{0}(k)}\,dk\,,
  \label{eq:gesc_main_R}\\
  \mathcal{A}_{\mathrm{M}}(\theta)
  &=
  \frac{
    \textstyle\int_{k>0}
    \abs{p_{\theta}(k)-p_{\theta}(-k)}\,dk
  }{
    \textstyle\int_{k>0}
    \left[p_{\theta}(k)+p_{\theta}(-k)\right]\,dk
  }\,,
  \label{eq:gesc_main_AM}\\
  \bar{k}_\theta
  &=
  \textstyle\int k p_{\theta}(k)\,dk\,,
  \label{eq:gesc_main_kbar}\\
  \sigma_k(\theta) 
  &= \sqrt{\textstyle\int 
    \left(k-\bar{k}_\theta\right)^2
    p_{\theta}(k)\,dk}\,.
  \label{eq:gesc_main_sigma}
\end{align}

All momentum integrals span the first Brillouin zone. $R$ quantifies redistribution relative to $\theta=0$, including symmetric and asymmetric changes, and $\mathcal{A}_{\mathrm M}$ captures inversion breaking. The mean momentum $\bar{k}$ quantifies net displacement of spectral weight and can remain small when shifts on opposite sides of $k=0$ cancel, while $\sigma_k$ characterizes the distribution width.

\begin{figure}[t]
\centering
\includegraphics{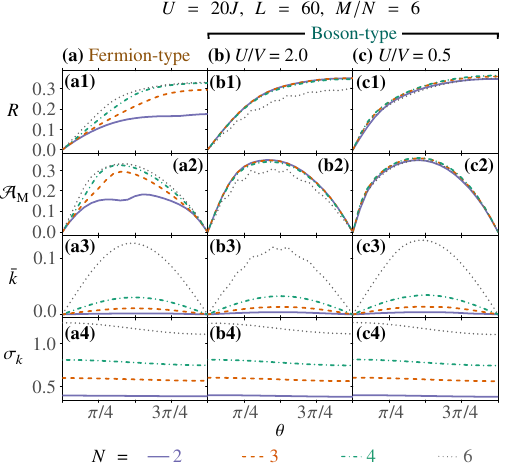}
\caption{\label{fig:gesc_main_nk_metrics}\textbf{Momentum reconstruction persists after inversion symmetry is restored.}
Four diagnostics quantify the intrinsic anyonic momentum distributions in \hyperref[fig:gesc_main_nk_heatmaps]{Fig.~\ref{fig:gesc_main_nk_heatmaps}}, using unit-normalized distributions over the full first Brillouin zone [Eqs.~\eqref{eq:gesc_main_R}--\eqref{eq:gesc_main_sigma}]. Redistribution $R$ remains finite at $\theta=\pi$, where the mirror asymmetry $\mathcal A_{\mathrm M}$ and mean momentum $\bar{k}$ vanish. The width $\sigma_k$ varies only weakly with $\theta$.\vspace{18pt}}
\end{figure}

\begin{figure}[t]
\centering
\includegraphics{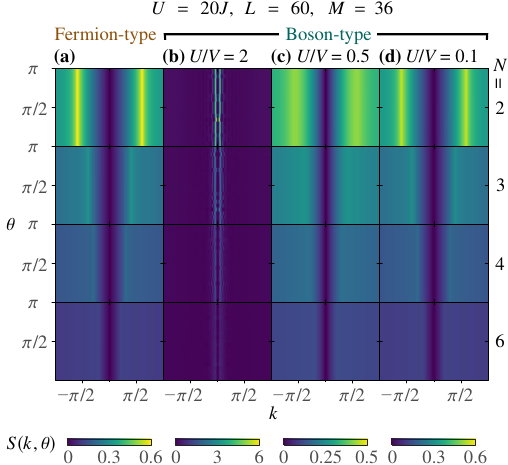}
\caption{\label{fig:gesc_main_Sk_heatmaps}\textbf{Interaction anisotropy controls the statistical sensitivity of spin correlations.} GESC spin structure factors $S(k,\theta)$ at fixed total population $M$. Boson-type anyons with $U/V=2$ concentrate spin correlations at small nonzero momenta, whereas the other cases exhibit broader antiferromagnetic profiles. Unlike Figs.~\ref{fig:gesc_main_nk_heatmaps} and \ref{fig:gesc_main_nk_metrics}, here the total filling is fixed at $M/L=0.6$.\vspace{13pt}}
\end{figure}

\hyperref[fig:gesc_main_nk_metrics]{Figure~\ref{fig:gesc_main_nk_metrics}} shows that fractional exchange statistics primarily redistributes spectral weight rather than strongly changing the overall momentum width. The redistribution $R$ grows with $\theta$ and remains finite at $\theta=\pi$, even though both $\mathcal{A}_{\mathrm M}$ and $\bar{k}$ return to zero. Restoring reflection symmetry therefore does not restore the initial distribution. Mirror asymmetry peaks at intermediate statistical phases $\theta/\pi\simeq0.40$--$0.45$. Increasing $N$ generally enhances redistribution and asymmetry for fermion-type anyons, whereas the component dependence is weaker for boson-type anyons. The mean momentum $\bar{k}$ increases strongly with $N$ and varies nonmonotonically with $\theta$ for both fermion- and boson-type anyons. In contrast, the width $\sigma_k$ also increases with $N$, but depends only weakly on $\theta$.

\subsubsection{Spin structure factor}
\label{sec:gesc_main_spin}
\hyperref[fig:gesc_main_Sk_heatmaps]{Figure~\ref{fig:gesc_main_Sk_heatmaps}} presents the spin structure factor at fixed $M=36$ and filling $M/L=0.6$. For fermion-type anyons \hyperref[fig:gesc_main_Sk_heatmaps]{[Fig.~\ref{fig:gesc_main_Sk_heatmaps}(a)]}, the central minimum, approximately linear rise at small nonzero $\abs{k}$, and symmetric peaks near $k=\pm2k_{\mathrm F}$ are characteristic of antiferromagnetic ordering. Since $k_{\mathrm F}=\pi M/(NL)$, increasing $N$ reduces the population per component and moves the peaks towards the origin, while their heights also decrease. The profiles remain essentially unchanged with $\theta$.

Boson-type anyons at $U/V=0.5$ \hyperref[fig:gesc_main_Sk_heatmaps]{[Fig.~\ref{fig:gesc_main_Sk_heatmaps}(c)]} show similar behavior, although with less sharp finite-momentum peaks. Reducing the ratio to $U/V=0.1$ \hyperref[fig:gesc_main_Sk_heatmaps]{[Fig.~\ref{fig:gesc_main_Sk_heatmaps}(d)]} brings the profiles close to their fermion-type counterparts, consistent with suppressed same-component virtual double occupancy as $V$ approaches the hard-core limit. Both ratios likewise show negligible statistical dependence.

In contrast, $U/V=2$ \hyperref[fig:gesc_main_Sk_heatmaps]{[Fig.~\ref{fig:gesc_main_Sk_heatmaps}(b)]} favors same-component alignment. Without the population constraint, a fully polarized ground state would exhibit a zero-momentum peak; imposed spin balance instead enforces $S(0)=0$. Same-component domains can accommodate the ferromagnetic tendency while preserving equal component populations. In this picture, the positive correlations within domains are neutralized by the negative correlations between domains of different components. The enhancement at small nonzero momenta is thus consistent with ferromagnetic ordering under the population constraint. The spin structure factor generally remains nearly independent of the statistical phase, with any residual dependence likely arising from domain-wall contributions to the spin correlations.

\section{\label{sec:statistical_origin}Statistical signatures}
Fractional exchange statistics can enter an observable through two distinct channels. First, the many-body state itself can depend on $\theta$; we refer to this as \textit{state-mediated} statistical dependence. Second, the observable may explicitly contain the nonlocal anyonic string generated by exchanging a particle through the intervening particles; we refer to this as \textit{operator-induced} statistical dependence. Appendix~\ref{app:gesc_statemediated_analysis} shows that, in the strong-interaction regime, the ground-state spin wave function depends only weakly on $\theta$: the coefficients $\mathcal C_\ell(\theta)$ predominantly undergo an overall rescaling rather than a reconstruction of their spatial profile. Consequently, pronounced ground-state signatures of fractional statistics are expected primarily in observables that explicitly retain the anyonic string.

\subsection{\label{sec:statistical_origin_operator}Operator-induced statistical dependence}
The GESC formalism provides a unique spin--charge perspective on the one-body density correlation, making the state- and operator-induced contributions particularly transparent. For $n>m$, we may write
\begin{align}
    C(m,n) = \sum_{m_p,n_p} D^{m,n}_{m_p,n_p}\, \omega^{\mathrm A}(m_p,n_p;\theta)\,,
\end{align}
where $D^{m,n}_{m_p,n_p}$ contains the charge contribution. Defining
$\zeta=n_p-m_p$, the phase-stripped permutation weights are
{\medmuskip=0mu
\thinmuskip=0mu
\begin{align}
        \omega^{\mathrm{F\lor B}}(m_p&,n_p;\theta) 
        = \frac{(\mp1)^{\zeta}}{N}
        \expval{ \mathcal E_{m_p,m_p+1}\ldots \mathcal E_{n_p-1,n_p} }{\chi(\theta)}\,,
\end{align}
}where $-1$ and $+1$ are for fermion- and boson-type anyons, respectively. The intrinsic anyonic correlation contains the additional statistical string,
\begin{align}
\omega^{\mathrm A}(m_p,n_p;\theta) = e^{-i\theta\zeta}\, \omega^{\mathrm{F\lor B}}(m_p,n_p;\theta)\,,
\end{align}
with the conjugate phase for the opposite propagation direction. Thus, $\omega^{\mathrm{F \lor B}}$ isolates the statistical dependence carried by the spin state, whereas $\omega^{\mathrm A}$ contains both the state-mediated contribution and the explicit statistical phase of the observable.

\begin{figure}
\centering
\includegraphics{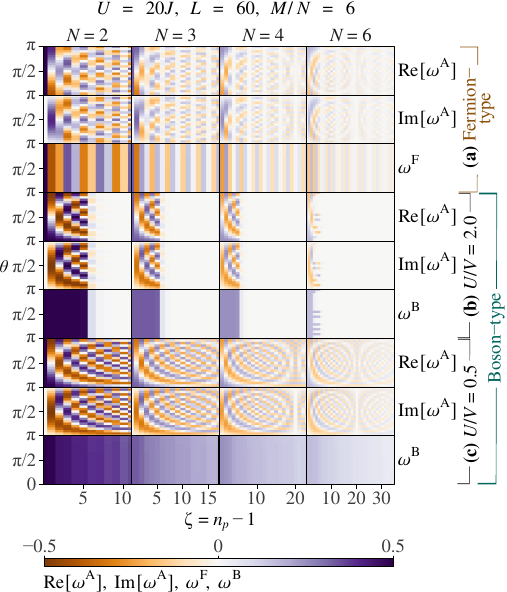}
\caption{\label{fig:spinexchange}\textbf{Explicit anyonic phases dominate the statistical dependence of correlations.} Permutation contributions to the one-body density correlations with ($\omega^{\mathrm A}$) and without ($\omega^{\mathrm{F \lor B}}$) the explicit anyonic phase, versus ordered-particle separation $\zeta=\abs{n_p-m_p}$, with $m_p=1$. The near collapse of the phase-stripped $\omega^{\mathrm{F \lor B}}$, contrasted with the strongly $\theta$-dependent $\omega^{\mathrm{A}}$, shows that the explicit statistical character of the observable provides the dominant ground-state $\theta$-dependence in the strong-interaction regime.}
\end{figure}

\hyperref[fig:spinexchange]{Figure~\ref{fig:spinexchange}} demonstrates this distinction. For fermion-type anyons \hyperref[fig:spinexchange]{[Fig.~\ref{fig:spinexchange}(row 3)]}, $\omega^{\mathrm F}$ shows the alternating sign structure associated with the fermionic permutation parity but remains nearly $\theta$ independent. This phase-stripped contribution enters the physical-fermion momentum distribution of Ref.~\cite{basak_anyon1}; its comparison with the intrinsic distribution in Figs.~2 and 5 and Appendix~D therein provides the full-model separation of state- and operator-induced statistical dependence. 

For boson-type anyons, $\omega^{\mathrm B}$ is likewise nearly $\theta$ independent, although its spatial structure reflects the different spin ground states. At $U/V=0.5$ \hyperref[fig:spinexchange]{[Fig.~\ref{fig:spinexchange}(row 9)]}, the permutation overlap $\langle\mathcal E_{m_p,m_p+1}\ldots \mathcal E_{n_p-1,n_p}\rangle$ remains positive and decreases smoothly with ordered-particle separation, whereas at $U/V=2$ \hyperref[fig:spinexchange]{[Fig.~\ref{fig:spinexchange}(row 6)]} the sharper boundaries are consistent with the same-component domain structure favored in the spin-balanced ground state.

The strong $\theta$ dependence appears only after the explicit factor $e^{-i\theta\zeta}$ is restored. It rotates the phase-stripped permutation weight in the complex plane, generating the pronounced structures in both $\Re[\omega^{\mathrm A}]$ \hyperref[fig:spinexchange]{[Fig.~\ref{fig:spinexchange}(rows 1, 4, 7)]} and $\Im[\omega^{\mathrm A}]$ \hyperref[fig:spinexchange]{[Fig.~\ref{fig:spinexchange}(rows 2, 5, 8)]}. Longer exchange strings accumulate the phase $\theta\zeta$ more rapidly. The imaginary contribution vanishes at $\theta=0$ and $\pi$ and becomes finite at intermediate statistics. This imaginary component is the real-space origin of the inversion asymmetry of the intrinsic momentum distribution. At $\theta=\pi$, the statistical string reduces to the parity factor $(-1)^\zeta$. The reduced amplitude of $\omega^{\mathrm A}$ and $\omega^{\mathrm{F\lor B}}$ with increasing $N$ largely reflects the explicit $1/N$ normalization of the permutation weight, rather than a suppression of the underlying permutation correlations.

The intrinsic momentum distribution is therefore a particularly sensitive probe because its one-body operator explicitly transfers a particle across a finite interval and carries the corresponding statistical string. More generally, off-diagonal observables whose operators explicitly contain the anyonic string are expected to exhibit pronounced statistical signatures in the ground state. Non-locality alone does not guarantee a strong statistical signature: the phase-stripped correlators are nonlocal but nearly independent of the statistical phase in the strong-interaction regime. The dominant dependence comes from the explicit anyonic string.

\subsection{\label{sec:statistical_origin_state}State-mediated statistical dependence}
Observables constructed only from local densities, such as the spin structure factor, contain no explicit anyonic string. Their ground-state statistical dependence must therefore originate entirely from that of the many-body state. As shown in Appendix~\ref{app:gesc_statemediated_analysis}, this contribution is weak in the strong-interaction ground state. Fractional exchange statistics rescales the bond-dependent spin-exchange coefficients without changing their spatial profile,
\begin{align}
\mathcal C_\ell(\theta) =
\mathcal C_\ell^{(0)}
+
2\mathcal C_\ell^{(+)}\cos\theta\,,
\end{align}
where $\mathcal C_\ell^{(0)}$ is the $\theta$-independent two-site contribution and $\mathcal C_\ell^{(+)}$ is the phase-stripped three-site contribution in Eq.~\eqref{eqn:CLdecop}; the corresponding term in Eq.~\eqref{eqn:HSCF1} carries the statistical phase $e^{i\theta}$. This is the direct outcome of $\mathcal C_\ell^{(+)}$ being purely real and results in the nearly $\theta$-independent spin structure factors above.

Dynamics provides a different route to amplify state-mediated statistical dependence.
The symmetric $\cos\theta$ contribution persists during the evolution, but the evolving charge state can generate a finite $\Im[\mathcal C_\ell^{(+)}(0,t)]$ and hence an additional antisymmetric contribution $\propto\sin\theta$. Appendix~\ref{app:gesc_statemediated_analysis} shows, from a short-time expansion, that this asymmetry is generated dynamically by the directed three-site charge-correlation terms. The symmetric and antisymmetric parts are
\begin{align}
\begin{split}
    \mathcal C^{(s)}_\ell(\theta,t) =
 &\;\mathcal C_\ell^{(0)}(0)
+
2\mathcal C_\ell^{(+)}(0,0)\cos\theta
\\&-
\frac{t^2}{2}
\expval{
\comm{H_{\mathrm C}}{
\comm{H_{\mathrm C}}{\hat{\mathcal C}_\ell^{(0)}}}
}{\varphi(0)}
\\&-t^2\;
\expval{
\comm{H_{\mathrm C}}{
\comm{H_{\mathrm C}}{\hat{\mathcal C}_\ell^{(+)}}}
}{\varphi(0)}\cos\theta,
\\
\mathcal C^{(a)}_\ell(\theta,t)
=
&-2t\,
\expval{
\comm{H_{\mathrm C}}{\hat{\mathcal C}_\ell^{(+)}}
}{\varphi(0)}\sin\theta.
\end{split}
\end{align}
The resulting $\mathcal C_\ell(\theta,t)$ coefficient thus exhibits both a statistics-dependent reduction of the exchange scale and, at intermediate statistics, a right--left asymmetry absent in the ground-state profile (\hyperref[fig:dynclcoeff]{Fig.~\ref{fig:dynclcoeff}}).

Although a local observable contains no explicit anyonic string, its dynamics can inherit statistical dependence from the evolving many-body state. This is particularly transparent for single impurity expansion: each spin exchange moves it one spin-chain site, so propagation samples an increasing sequence of bond-dependent coefficients $\mathcal C_\ell(\theta,t)$. Their statistical dependence can accumulate over successive exchanges, producing pronounced $\theta$ dependence in the local impurity density despite the ground-state expectation value being nearly $\theta$ independent. This distinction motivates the transport analysis in Sec.~\ref{sec:dynamics}: static ground-state signatures are most pronounced when the observable itself contains the statistical string, whereas time evolution can amplify the weaker state-mediated dependence and render its impact tangible in local density and transport observables.

\section{\label{sec:dynamics}Dynamical applications}
We contrast dipole oscillations, governed by \textit{charge motion}, with impurity release, which probes \textit{spin exchange}. These protocols connect the GESC to the Anyon--Hubbard transport studied in Ref.~\cite{basak_anyon1}. We then extend the impurity problem to two particles, considering both identical and distinguishable impurities. For boson-type anyons, these two preparations probe the different roles of intra- and intercomponent interactions and their interplay with fractional exchange statistics, allowing us to examine how particle identity and interaction anisotropy reshape the exchange-mediated transport.

\subsection{\label{sec:dipole}Dipole oscillations}
Dipole oscillations are generated by a sudden displacement $\delta$ of the harmonic-trap center and are characterized by the center-of-mass motion $ x_{\mathrm{com}}(t) =\sum_i x_i\expval{n_i(t)}/M$ with $x_i$ denoting the position of site $i$ from the center of the system. Since $x_{\mathrm{com}}$ is constructed entirely from the local density, it acts only on the charge degrees of freedom. In the strongly interacting GESC formalism, its evolution is therefore determined by the spinless-fermion charge Hamiltonian
{\medmuskip=0mu
\thinmuskip=0mu
\begin{align}
    H_{\mathrm C}(t) = -J\sum_i \left(
        f_i^\dagger f_{i+1} + f_{i+1}^\dagger f_i
    \right) + \sum_i \Omega\left[x_i-\delta\Theta(t)\right]^2 n_i\,,
\label{eqn:tc0}
\end{align}
}which has no dependence on $\theta$.

This can be seen directly from the Heisenberg equation of motion. Using
$
[n_i,f_j^\dagger f_k]
=
(\delta_{ij}-\delta_{ik})f_j^\dagger f_k
$,
the local density satisfies
$
\dot{n}_i
=
i[H_{\mathrm C},n_i]
=
\mathcal F_{i-1,i}
-
\mathcal F_{i,i+1}
$
where
$\mathcal F_{i,i+1}
=
-iJ
(
f_i^\dagger f_{i+1}
-
f_{i+1}^\dagger f_i
)$.
Consequently,
$\dot{x}_{\mathrm{com}}
=
{M}^{-1}
\sum_i
(x_{i+1}-x_i)
\expval{\mathcal F_{i,i+1}}$
is governed entirely by the $\theta$-independent charge evolution. Since the initial charge wave function is likewise independent of $\theta$, the GESC formalism predicts identical dipole oscillations for all $\theta$ in the strong-interaction regime.

This directly contrasts with the impurity release dynamics below. Dipole motion redistributes the charge without requiring a rearrangement of the ordered spin sequence and therefore does not sample the $\theta$-dependent spin-exchange coefficients. The weak residual statistical dependence of the center-of-mass motion in the strong-interaction calculations of Ref.~\cite{basak_anyon1} (Sec.~IV~A and Fig.~6 therein) reflects finite-interaction corrections beyond the $\theta$-independent leading-order charge dynamics. Stronger interactions suppress these corrections, driving the dynamics toward $\theta$-independent GESC behavior.

\subsection{\label{sec:impurity_release}Impurity release}
Here, we apply the GESC formalism to impurity-release dynamics, where the role of statistics enters through the interplay between charge evolution and spin exchange.

\subsubsection{Preparation and effective Hamiltonians}
We consider an anyonic host gas formed by one internal component and immerse either one or two impurities belonging to other component(s). The impurity particles are initially localized at the center of the lattice by a local pinning field $h_z(t)$. At $t=0$, the pinning is removed while the harmonic confinement remains, and we monitor the subsequent impurity density as a function of site and time.

Within GESC, the spatial preparation and subsequent charge evolution are described by
{\medmuskip=1mu
\thinmuskip=1mu
\begin{align}
H_{\mathrm C}(t) = -J\sum_{\langle i,j\rangle} \left( f_i^\dagger f_j +\mathrm{H.c.} \right)
        +\Omega\sum_i x_i^2 n_i - h_z(t)\sum_{i\in\mathbb C}n_i\,,
\label{eqn:impurity_charge}
\end{align}
}where $\mathbb C$ denotes the central pinning site(s) and $h_z(t) = h_z\Theta(-t)$. 

The localized potential ensures the charge ground state pins particles at the desired central site(s) prior to the quench. Since the charge Hamiltonian contains no information about the internal component(s), the impurity identity is imposed separately through a component-selective term in the GESC Hamiltonian. For a pinning site $i_p$, its projection onto the ordered-particle chain is determined by
\begin{align}
    \mathcal D_{\ell}^{(i_p)} = \expval{
    \delta_{{\textstyle\sum\nolimits_{i=1}^{i_p-1}}
{\scriptstyle\overline{n}_i,\ell-1}} n_{i_p} }{\varphi}\,,
\end{align}
which determines the probability the particle occupying physical site $i_p$ corresponds to the $\ell$th ordered particle. Throughout this section we set $h_z = 8J$ and $\Omega = 5\times10^{-3}J$. 

The spin evolution is governed by the corresponding fermion- or boson-type GESC Hamiltonian [Eqs.~\eqref{eqn:HSCF} and \eqref{eqn:HSCB}], with the prescription of an appropriate component-selective pinning term. At $t=0$, all pinning terms are removed and the subsequent spin dynamics are governed by the time-dependent $\mathcal C_\ell(\theta,t)$. The exact form of the GESC Hamiltonian depends on whether the impurities are fermion- or boson-type and, for the latter, on the intra- and intercomponent interactions; we therefore give the relevant form together with each impurity configuration in what follows.

\subsubsection{Exchange coefficients and dynamical observables}

For a given charge preparation and evolution, the coefficients $\mathcal C_\ell(\theta,t)$ are fixed entirely by the charge wave function. They are therefore identical for fermion- and boson-type anyons, and independent of interaction strength and the number of components, provided the system size, the number of particles, the trap parameters are the same. The distinction between these cases enters through the spin operators and interaction prefactors in the corresponding GESC Hamiltonians. Different impurity configurations, such as the single- and two-impurity systems that we will consider, generally have different charge preparations and therefore their own $\mathcal C_\ell(\theta,t)$.

As shown in Appendix~\ref{app:gesc_statemediated_analysis}, the dynamical coefficients differ from their ground-state counterparts: their magnitude decreases with increasing $\theta$, while intermediate statistical phases generate right--left asymmetry in their spatiotemporal profiles. Since the spin-exchange energy scales as $J^2\mathcal C_\ell(\theta,t)/U$, these variations provide the microscopic link between fractional statistics and the impurity dynamics discussed below.

To characterize the impurity motion uniformly across the different configurations, we define the total impurity density
$
n_{i,\mathrm{imp}}(t)
=
\textstyle\sum_{\alpha\in\mathbb I}
\expval{n_{i,\alpha}(t)}
$,
where $\mathbb I$ denotes the impurity component(s).
The normalized impurity distribution is
$
p_i(t)
=
n_{i,\mathrm{imp}}(t)/
\textstyle\sum_j n_{j,\mathrm{imp}}(t)
$.
Thus, $p_i(t)=\expval{n_{i,\downarrow}(t)}$ for a single
impurity and $p_i(t)=n_{i,\mathrm{imp}}(t)/2$ for two impurities.
We characterize the spatial extent relative to the fixed release
center $i_{\mathrm c}$ by the root-mean-square (RMS) distance
$
W(t)=\sqrt{\textstyle\sum_i(i-i_{\mathrm c})^2p_i(t)}
$,
and the directional asymmetry by the population imbalance
$
\mathcal A(t)
=
\textstyle\sum_{i>i_{\mathrm c}}p_i(t)
-
\textstyle\sum_{i<i_{\mathrm c}}p_i(t)
$.

\subsubsection{Single-impurity dynamics}

We consider a two-component gas with $M^\uparrow=M-1=10$ and a single spin-down impurity, $M^\downarrow=1$, for $L=21$ and $U=40J$. The fermion-type preparation follows the single-impurity release protocol of Ref.~\cite{basak_anyon1} (Sec.~IV~B and Fig.~7 therein), with $M=11$, $U=40J$, and $h_z=8J$ here, compared with $M=10$, $U=20J$, and $h_z=4J$ in its strong-interaction example. A longitudinal field initially pins the impurity at the central physical site $i_{\mathrm c}$. At $t=0$ the field is removed, while the harmonic confinement remains. In Eq.~\eqref{eqn:impurity_charge}, this corresponds to $\mathbb C=\{i_{\mathrm c}\}$. The corresponding spin Hamiltonians are
{\medmuskip=0mu
\thinmuskip=0mu
\begin{align}
    H^{\mathrm{S:F}}(t)
    ={}&-\dfrac{J^2}{U}
    \sum_{\ell}\mathcal{C}_\ell(\theta,t)
    \left(I-\mathcal{E}_{\ell,\ell+1}\right)
    +h_z(t)\sum_{\ell}\mathcal{D}^{(i_{\mathrm c})}_{\ell}(t)\sigma^z_{\ell}\,,
    \label{eqn:tsc1}\\
    H^{\mathrm{S:B}}(t)
    ={}&-\frac{J^2}{U}\sum_{\ell}\mathcal C_\ell(\theta,t)
    \left(I+\mathcal E_{\ell,\ell+1}\right) -J^2\left(\frac{1}{V}-\frac{1}{U}\right)
    \sum_{\ell} \bigg[ \mathcal C_\ell(\theta,t)\nonumber\\
    &\qquad\times
    \left(I+\sigma_{\ell}^z\sigma_{\ell+1}^z\right) \bigg]+h_z(t)\sum_{\ell}\mathcal{D}^{(i_{\mathrm c})}_{\ell}(t)\sigma_{\ell}^z\,.
    \label{eqn:tsc1B}
\end{align}
}For the initial preparation, we include the pinning field in the projected charge and spin Hamiltonians while neglecting its corrections to the virtual-state energy denominators. All pinning terms vanish after release. Charge-sector pinning localizes a particle at $i_{\mathrm c}$, while the longitudinal term in the spin Hamiltonian selects that particle to be the impurity component. The fermion-type GESC dynamics agree well with the full Anyon--Hubbard evolution, as benchmarked in Appendix~\ref{app:gesc_dynamics_benchmarks}.

\begin{figure}[t]
\centering
\includegraphics{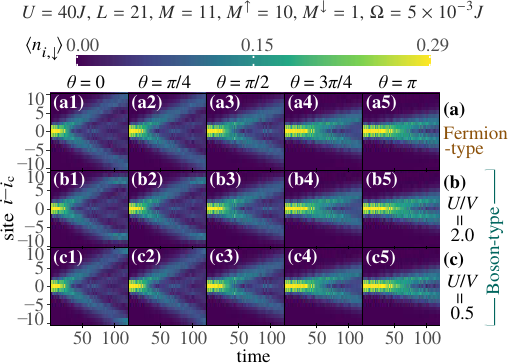}
\caption{\label{fig:denspinchain}\textbf{Fractional exchange statistics tunes the extent and directionality of impurity propagation.} GESC time evolution, following release of a single centrally pinned impurity, of the physical-lattice impurity density $\langle n_i^\downarrow(t)\rangle$. The initial pinning field $h_z=8J$ is removed at $t=0$. The narrowing propagation fronts track the reduction in effective exchange. Colors saturate at the 99th percentile. Time is expressed in units of $J^{-1}$.}
\end{figure}

\hyperref[fig:denspinchain]{Figure~\ref{fig:denspinchain}} shows single-impurity time evolution for fermion- and boson-type anyons. In all three cases, increasing $\theta$ progressively suppresses the broad propagation seen at $\theta=0$, with increasingly localized population near the release site. Intermediate statistical phases also produce weak right--left asymmetry. The origin
of these trends follows directly from the properties of
$\mathcal C_\ell(\theta,t)$ established in
Sec.~\ref{sec:statistical_origin_state} and
Appendix~\ref{app:gesc_statemediated_analysis}.

For a fermion-type impurity localized at spin-chain site $\ell_0$ at time
$t_0$, the short-time expansion in Appendix~\ref{app:gesc_statemediated_analysis} gives
\begin{align}
\begin{split}
\expval{n_{\ell_0-1,\downarrow}(t_0+\delta t)}
&=
\tfrac{J^4}{U^2}
\mathcal C_{\ell_0-1}^2(\theta,t_0)\delta t^2
+\order{\delta t^3},
\\
\expval{n_{\ell_0+1,\downarrow}(t_0+\delta t)}
&=
\tfrac{J^4}{U^2}
\mathcal C_{\ell_0}^2(\theta,t_0)\delta t^2
+\order{\delta t^3}.
\label{eqn:single_impurity_short_main}
\end{split}
\end{align}
Thus, the reduction in the magnitude of $\mathcal C_\ell(\theta,t)$ with increasing $\theta$ directly suppresses the transfer of the impurity between neighboring ordered-particle sites. Similarly, the directional impurity dynamics follows from the spatial asymmetry of $\mathcal C_\ell(\theta,t)$: unequal $\mathcal C_{\ell_0-1}$ and $\mathcal C_{\ell_0}$ produce unequal short-time transfer to the left and right neighboring spin-chain sites. Reaching more distant sites requires more number of successive exchanges and thus samples an increasing sequence of these coefficients. Their reduced magnitudes compound during propagation, producing the progressively narrower fronts in Fig.~\ref{fig:denspinchain}, while their spatial asymmetry is sampled over additional bonds, allowing the right--left imbalance to build up and produce the weak inversion asymmetry at intermediate statistical phases.

Boson-type dynamics exhibit the same statistics-induced suppression, with the extent of propagation additionally controlled by $U/V$. At $U/V=2$, the spin Hamiltonian favors same-component alignment, further inhibiting impurity motion, which requires exchanges with the surrounding majority component. At $U/V=0.5$, antiferromagnetic component configurations are more favorable, allowing the impurity to propagate farther. Since both cases share the same charge evolution and $\mathcal C_\ell(\theta,t)$, this difference originates in the $V$-dependent spin Hamiltonian rather than from the charge dynamics.

\begin{figure}[t]
\centering
\includegraphics{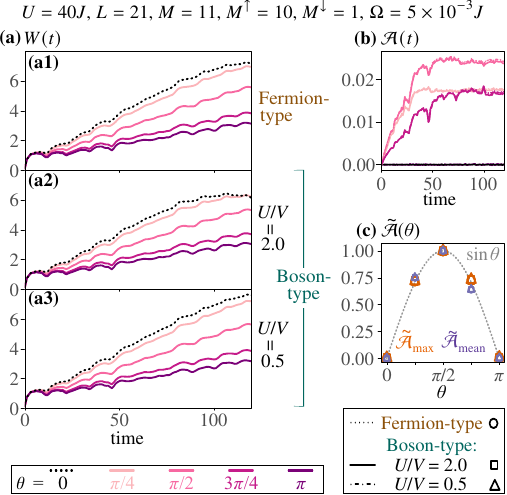}
\caption{\label{fig:impuritymetrics}\textbf{Interaction tuning modifies statistics-dependent impurity propagation more strongly than the directional imbalance.} Physical-lattice (a) RMS displacement $W(t)$ and (b) right--left population imbalance $\mathcal A(t)$, together with (c) normalized maximum and time-averaged magnitudes of the directional asymmetry. The near-zero imbalance at $\theta \in \{0,\;\pi\}$ and its peak near $\theta=\pi/2$ are consistent with the direction-dependent statistical phases $e^{\pm i\theta}$ in occupied-neighbor-assisted three-site tunneling. Parameters and release protocol are as in \hyperref[fig:denspinchain]{Fig.~\ref{fig:denspinchain}}. Time is expressed in units of $J^{-1}$.}
\end{figure}

\hyperref[fig:impuritymetrics]{Figure~\ref{fig:impuritymetrics}(a)} quantifies the suppression of impurity propagation through the RMS displacement $W(t)$. For both anyon types, increasing $\theta$ generally reduces the growth of $W(t)$, consistent with the reduced magnitude of $\mathcal C_\ell(\theta,t)$  discussed in Appendix~\ref{app:gesc_statemediated_analysis}. For boson-type anyons, $U/V=2$ produces smaller displacements than $U/V=0.5$, especially near $\theta=0$.

The directional imbalance exhibits a different statistical dependence \hyperref[fig:impuritymetrics]{[Fig.~\ref{fig:impuritymetrics}(b)]}. It remains negligible at $\theta=0$ and $\pi$, reaching its largest value at $\theta=\pi/2$. At intermediate phases, $\mathcal A(t)$ initially rises, then fluctuates around a nearly constant positive value while $W(t)$ continues to grow. Unlike the displacement, the imbalance curves nearly coincide for fermion-type anyons and both bosonic interaction ratios. Thus, the interaction ratio controls how far the impurity propagates much more strongly than the right--left population imbalance.

The normalized maximum asymmetry closely follows the $\sin\theta$ reference closely \hyperref[fig:impuritymetrics]{[Fig.~\ref{fig:impuritymetrics}(c)]}, which is consistent with the antisymmetric contribution to $\mathcal C_\ell(\theta,t)$ derived in Appendix~\ref{app:gesc_statemediated_analysis}. It vanishes at $\theta=0$ and $\pi$ and has its largest statistical prefactor at $\theta=\pi/2$. The short-time transfers in Eq.~\eqref{eqn:single_impurity_short_main} show that this exchange asymmetry produces unequal transfer to the two neighboring spin-chain sites, resulting in a right--left difference that can accumulate over successive exchanges (see Appendix~\ref{app:gesc_statemediated_analysis}). Its similar dependence across the three cases is also consistent with their shared $\mathcal C_\ell(\theta,t)$. The time-averaged asymmetry deviates more from $\sin\theta$ because it also depends on how quickly the imbalance develops. For fermion-type anyons, the vanishing imbalance at $\theta=0$ and $\pi$ agrees with the symmetry analysis in Appendix~F of Ref.~\cite{basak_anyon1}, which derives the inversion relation in the full Anyon--Hubbard model.

\subsubsection{Two-impurity dynamics}
We next consider \textit{two} impurities embedded in a single-component host gas, initially pinned near the lattice center and released simultaneously. For both anyon types, we compare identical impurities of the same component with distinguishable impurities of different components. This comparison probes how impurity identity and, for boson-type anyons, interaction anisotropy shape the $\theta$-dependent expansion.

\begin{figure*}
\centering
\includegraphics{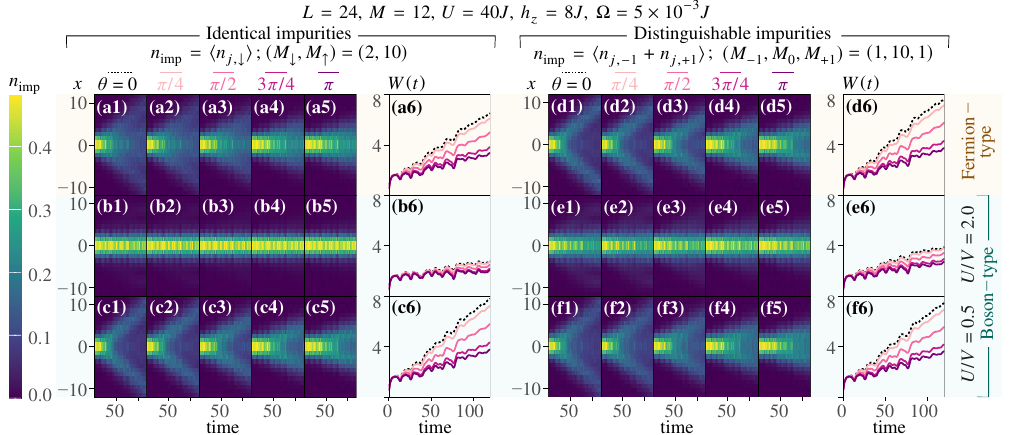}
\caption{\label{fig:two_impurity_comparison}\textbf{Statistical control of two-impurity expansion.}
Total impurity density $n_{\mathrm{imp}}(j,t)$ (columns 1--5) and RMS distance from the release center $W(t)$ (column 6) for identical impurities (left block; a--c) and distinguishable impurities (right block; d--f). Both preparations exhibit suppressed propagation as $\theta$ approaches $\pi$, with the strongest confinement for boson-type anyons at $U/V=2.0$. The impurities are initially pinned near sites $j=12,13$ by a field $h_z=8J$, which is removed at $t=0$. Position is measured relative to the release center, $x=j-12.5$, in lattice spacings. Colors saturate at the 99th percentile. Time is expressed in units of $J^{-1}$.}
\end{figure*}

\paragraph{Impurities in the same component.}
We consider a two-component gas with $M^\uparrow=M-2=10$ and two spin-down impurities, $M^\downarrow=2$, for $L=24$ and $U=40J$. A longitudinal field initially pins the impurities near the central physical sites $i_{\mathrm c,1}=12$ and $i_{\mathrm c,2}=13$. At $t=0$ the field is removed, while the harmonic confinement remains. In Eq.~\eqref{eqn:impurity_charge}, this corresponds to $\mathbb C =\{i_{\mathrm c,1},i_{\mathrm c,2}\}$. The corresponding spin Hamiltonians are
\begin{align}
    H^{\mathrm{S:F}}(t)
    = &-\dfrac{J^2}{U}
    \sum_{\ell}\mathcal{C}_\ell(\theta,t)
    \left(I-\mathcal{E}_{\ell,\ell+1}\right) \notag\\&+h_z(t)\sum_{\ell}\left[\mathcal{D}^{(i_{\mathrm c,1})}_{\ell}(t) + \mathcal{D}^{(i_{\mathrm c,2})}_{\ell}(t)\right]\sigma^z_{\ell}\,,
    \label{eqn:tsc2}\\
    H^{\mathrm{S:B}}(t)
    = &-\frac{J^2}{U}\sum_{\ell}\mathcal C_\ell(\theta,t)
    \left(I+\mathcal E_{\ell,\ell+1}\right) \notag\\&-J^2\left(\frac{1}{V}-\frac{1}{U}\right)
    \sum_{\ell}\mathcal C_\ell(\theta,t)
    \left(I+\sigma_{\ell}^z\sigma_{\ell+1}^z\right) \notag\\&+h_z(t)\sum_{\ell}\left[\mathcal{D}^{(i_{\mathrm c,1})}_{\ell}(t) + \mathcal{D}^{(i_{\mathrm c,2})}_{\ell}(t)\right]\sigma_{\ell}^z\,.
    \label{eqn:tsc2B}
\end{align}

\hyperref[fig:two_impurity_comparison]{Figure~\ref{fig:two_impurity_comparison}(a--c)} shows the statistical trends found for a single impurity persist for two identical impurities: propagation is broad at small $\theta$, while increasing $\theta$ suppresses expanding fronts and retains more population near the release region. The corresponding $W(t)$ curves show the same reduction of the spatial extent with increasing $\theta$. Thus, introducing a second impurity does not alter the mechanism identified above: impurity motion proceeds through successive spin exchanges with amplitudes set by the time-dependent $\mathcal C_\ell(\theta,t)$, allowing $\theta$ dependence to accumulate during propagation.

Boson-type anyons show a pronounced dependence on interaction anisotropy. At $U/V=0.5$, propagation resembles the fermion-type case and is progressively suppressed as $\theta$ increases. At $U/V=2$, a strong central population persists during the evolution, with substantially smaller $W(t)$. In this regime, the spin Hamiltonian favors same-component alignment; with two identical impurities, the same-component impurity pair can directly participate in this tendency, enhancing retention near the release region. This confinement therefore reflects the interplay between the statistics-dependent exchange amplitudes and the additional intracomponent interaction channel available to two identical boson-type impurities.

\paragraph{Impurities in different components.}
We consider a three-component gas with $M^0=M-2=10$ and the distinguishable impurities $M^{-1}= M^{1}=1$. All other parameters remain the same. The corresponding spin Hamiltonians are
{\medmuskip=0mu
\thinmuskip=0mu
\begin{align}
        &H^{\mathrm{S:F}}(t) \notag\\&=-\dfrac{J^2}{U} \sum_{\ell}\mathcal{C}_\ell(\theta,t)
        \left(I-\mathcal{E}_{\ell,\ell+1}\right) \notag\\&\quad+h_z(t)\sum_{\ell}\left[
            \mathcal{D}^{(i_{\mathrm c,1})}_{\ell}(t) + \mathcal{D}^{(i_{\mathrm c,2})}_{\ell}(t)
        \right]\left( S^z_{\ell,(-10)}-S^z_{\ell,(01)} \right)\,,
    \label{eqn:tsc3}\\
        &H^{\mathrm{S:B}}(t) \notag\\&= -\frac{J^2}{U}\smashoperator[r]{\sum_{\ell,\sigma<\sigma'}}\mathcal C_\ell
        (\theta,t) \left(I+\mathcal E_{\ell,\ell+1}\right)\hat{P}^{\sigma,\sigma'}_{\ell,\ell+1}
        \notag\\&\quad-J^2\left(\frac{1}{2V}-\frac{1}{U}\right) \smashoperator{\sum_{\mathrlap{\!\!\!\ell,\sigma<\sigma'}}}
        \mathcal C_\ell(\theta,t) \left(I+S_{\ell,(\sigma\sigma')}^zS_{\ell+1,(\sigma\sigma')}^z\right)
        \hat{P}^{\sigma,\sigma'}_{\ell,\ell+1}\notag \\
        &\quad+h_z(t)\sum_{\ell}\left[
            \mathcal{D}^{(i_{\mathrm c,1})}_{\ell}(t) + \mathcal{D}^{(i_{\mathrm c,2})}_{\ell}(t)
        \right]\left( S^z_{\ell,(-10)}-S^z_{\ell,(01)} \right)\,.
    \label{eqn:tsc3B}
\end{align}}

\hyperref[fig:two_impurity_comparison]{Figure~\ref{fig:two_impurity_comparison}(d--f)} shows that the statistics-induced suppression of propagation also persists for two distinguishable impurities, but their propagation differs from the identical-impurity case because the two impurities now belong to different components. Since both preparations have the same charge evolution, they share the same $\mathcal C_\ell(\theta,t)$; the differences therefore originate from the spin Hamiltonian rather than from the statistics-dependent charge coefficients. For fermion-type anyons, distinguishable impurities are displaced farther than the identical pair while retaining the same overall suppression with increasing $\theta$. For identical impurities, $(I-\mathcal E_{\ell,\ell+1})\ket{\downarrow\downarrow}=0$, whereas for distinguishable impurities $(I-\mathcal E_{\ell,\ell+1})\ket{-1,+1}\neq0$. This difference in the spin-exchange term provides a microscopic distinction between the two impurity cases. 

For boson-type anyons, the distinction between identical and distinguishable impurities is governed by interaction anisotropy. At $U/V=2$, the intracomponent spin-exchange coefficient is stronger than the intercomponent one and favors keeping identical impurities together, suppressing their propagation more strongly than for distinguishable impurities. At $U/V=0.5$, stronger intercomponent exchange reverses this: distinguishable impurities propagate less than the identical pair. Both preparations retain suppressed expansion with increasing $\theta$, with interaction anisotropy determining the overall difference between identical and distinguishable impurities.

Together, these comparisons separate the common statistical suppression from the effects of impurity identity. The former is inherited from the shared $\mathcal C_\ell(\theta,t)$, whose reduced magnitude suppresses successive spin exchanges, whereas the differences between identical and distinguishable impurities reflect the spin-sector exchange structure.

The broad range of ground-state and dynamical calculations presented here is made possible by the computational efficiency of the GESC formalism. As demonstrated in Appendix~\ref{app:gesc_efficiency}, separating the problem into a reusable charge calculation and an effective spin chain yields a substantial computational advantage over the full Anyon--Hubbard approach. These gains are particularly pronounced for scans over the statistical phase, since the charge solution is computed once and reused across values of $\theta$, allowing finely resolved studies of momentum-space reconstruction and impurity transport.

\section{\label{sec:summary}Summary and outlook}
We developed a generalized effective spin-chain (GESC) formalism for strongly interacting $N$-component anyons in a one-dimensional optical lattice, considering both fermion-type and boson-type exchange relations. Mapping particle motion onto spinless fermions and spin states onto an ordered chain provides both a computationally efficient description and a physically transparent spin--charge-separated picture of multicomponent anyonic matter. Benchmarking against full Anyon--Hubbard calculations shows that the GESC formalism captures the strong-interaction ground-state and nonequilibrium dynamics, while enabling access to larger system sizes, higher numbers of components, finely resolved statistical-phase scans, and substantially longer propagation times.

A key strength of the GESC formalism is that it also explicitly pinpoints how fractional exchange statistics enters the low-energy manifold. To leading order in the strong-interaction regime, the charge Hamiltonian reduces to that of spinless fermions and is independent of the statistical phase, which instead enters the spin sector through the bond-dependent spin-exchange coefficients $\mathcal{C}_\ell(\theta)$ generated by virtual tunneling through intermediate doubly occupied states. In particular, the statistical phases carried by occupied-neighbor-assisted three-site processes enter $\mathcal C_\ell(\theta)$. The GESC Hamiltonian thus delineates the contribution of fractional exchange statistics to the effective exchange physics even though real double occupancy is suppressed in the low-energy manifold.

For the ground-state properties, changing the statistical phase nearly uniformly rescales $\mathcal{C}_\ell(\theta)$, leaving the spin ground state largely unchanged. This explains why intrinsic momentum distributions retain pronounced statistical reconstruction through their explicit anyonic strings, while the spin structure factor depends weakly on statistics. For boson-type anyons, interaction anisotropy additionally selects distinct correlation regimes, ranging from antiferromagnetic ordering to same-component domains within the spin-balanced sector. These results distinguish statistical signatures carried by the state from those encoded explicitly in the observable.

Out of equilibrium, the spin--charge-separated representation distinguishes charge-dominated motion from exchange-mediated transport. Dipole oscillations are independent of statistics within leading-order charge dynamics; residual dependence in the full model reflects finite-interaction corrections. Impurity transport instead samples statistics-dependent $\mathcal{C}_\ell(\theta,t)$ throughout the time evolution. The microscopic virtual tunneling phases combine into real spin-exchange coefficients whose reduced magnitude suppresses propagation and whose spatial inversion-asymmetry produces directional imbalance. For fermion-type anyons, confinement is strongest toward the pseudobosonic endpoint. Thus, dynamics has potential to amplify statistical signatures in local impurity densities even when ground-state observables without an explicit anyonic string remain weakly dependent on the statistical phase.

Suppressed expansion persists for identical and distinguishable impurity pairs, while their relative propagation reflects the interplay between impurity identity and interaction anisotropy. For fermion-type anyons, distinguishable impurities propagate farther because they can virtually occupy the same site, a process forbidden by Pauli exclusion for identical impurities. For boson-type anyons, interaction anisotropy controls this difference and can even make identical impurities propagate farther. With identical charge preparations, both pairs share the same exchange coefficients; their different propagation therefore originates in the spin Hamiltonian. Impurity identity thus modifies transport without removal of statistical suppression.

\paragraph*{Future avenues.}
The GESC formalism provides quantitative guidance for experiments that use impurity dynamics to probe fractional exchange statistics through local density measurements. Comparing charge motion with spin transport, and varying impurity identity and interaction anisotropy, offers direct tests of how fractional exchange statistics survives strong interactions and controls many-body dynamics. Looking ahead, the scalability and physical transparency of the GESC formalism open directions difficult to access through direct Anyon--Hubbard calculations. Of particular interest are dynamical fermionization and bosonization, long-time relaxation and thermalization, statistics and interaction quenches, and transport with larger particle and component numbers. Component-dependent statistical phases and anisotropic interactions provide natural extensions in which qualitatively new effective spin couplings could emerge. The GESC formalism can be further generalized to more complex time-dependent potentials and driving protocols, enabling studies of strongly interacting spinor gases with arbitrary spin and fractional exchange statistics in regimes where the interplay of statistics, correlations, and nonequilibrium dynamics is expected to yield especially rich physics.

\begin{acknowledgments}
This study was supported by the National Science Foundation under Grant No.~PHY-2513089 and The Welch Foundation under Grant No.~C-1669.
SB acknowledges support by the National Science Foundation through Grant No.~PHY-2409311.
XWG acknowledges support from the National Natural Science Foundation of China (NSFC) Key Grants Nos.~U25D8013, 92365202, 12465001 and the Innovation Program for Quantum Science and Technology Grant Nos.~2021ZD0302000 and 2023ZD0300404.
This work was supported in part by the Big-Data Private-Cloud Research Cyberinfrastructure MRI-award funded by NSF under grant CNS-1338099 and by Rice University's Center for Research Computing (CRC). 
Some of the computing for this project was performed at the OU Supercomputing Center for Education \& Research (OSCER) at the University of Oklahoma (OU).
\end{acknowledgments}

\appendix
\let\addtocontents\origaddtocontents
\begin{center}
\PRLsep
\small\textbf{\\APPENDICES\vspace{-2em}}
\end{center}
\begingroup
\makeatletter
\renewcommand{\tocname}{}
\renewcommand{\baselinestretch}{1}\selectfont
\setlength{\parskip}{4pt}
\def\l@f@section{\addpenalty{\@secpenalty}}
\vspace{-4em}
\tableofcontents
\par
\makeatother
\endgroup

\section{\label{app:fermionic_derivation}GESC Hamiltonian for fermion-type anyons}
Consider a system of $N$-component anyons in a 1D optical lattice that anticommute on the same site and follow generalized commutation rules off site. Let the number of anyons be $M$ and the number of lattice sites be $L$. We derive the GESC Hamiltonian starting from the mapped Fermi--Hubbard Hamiltonian [Eq.~\eqref{eqn:fermiH}].

Define complementary subspaces $H_{\mathcal{Q}^0}$ (low-energy: site occupancy $\leq$ 1) with projector $\mathcal{Q}^0$ and $H_{\mathcal{Q}^1}$ (high-energy: occupancy $>$ 1) with projector $\mathcal{Q}^1(=1-\mathcal{Q}^0)$. A configuration belongs to the high-energy subspace if even a single site is multiply occupied. The effective Hamiltonian (up to second order) can be expressed as \cite{cazalilla_one-dimensional_2003}
\begin{align}
H_{\mathrm{eff}} = \mathcal{Q}^0 H_{\mathrm{F}} \mathcal{Q}^0-\mathcal{Q}^0 H_{\mathrm{F}}  \mathcal{Q}^1\dfrac{1}{\mathcal{Q}^1 H_{\mathrm{F}}  \mathcal{Q}^1}\mathcal{Q}^1H_{\mathrm{F}} \mathcal{Q}^0\,. \label{eqn:A1}
\end{align}
The zeroth-order term describes fermionic hopping:
{\medmuskip=0mu
\thinmuskip=0mu
\begin{align*}
\mathcal{Q}^0 H_{\mathrm{F}} \mathcal{Q}^0 = -J\sum\limits_{j,\alpha}\mathcal{Q}^0(c_{j,\alpha}^{\dagger}e^{-i\theta n_{j}}c_{j+1,\alpha}+c_{j+1,\alpha}^{\dagger}e^{+i\theta n_{j}}c_{j,\alpha})\mathcal{Q}^0
\end{align*}
}For an allowed hop within $\mathcal Q^0$, the density-dependent phase acts on the destination site before particle creation for a leftward hop, and on the departure site after particle annihilation for a rightward hop. In both cases, that site is empty when the phase acts, so the phase factor reduces to unity.

To enforce the constraint of at most one particle per site, we introduce the projected fermionic annihilation and creation operators,
\begin{align*}
    f_{j,\alpha} = c_{j,\alpha}\prod\limits_{\beta \neq \alpha}(1-n_{j,\beta}) \;\text{~and~}\;f_{j,\alpha}^{\dagger} = c_{j,\alpha}^{\dagger}\prod\limits_{\beta \neq \alpha}(1-n_{j,\beta})\,.
\end{align*}
The corresponding number operator is
\begin{align*}
    \bar{n}_{j,\alpha} =  n_{j,\alpha}\prod\limits_{\beta \neq \alpha}(1-n_{j,\beta})\,.
\end{align*}
In terms of these projected operators, we obtain
\begin{align}
\mathcal{Q}^0 H_{\mathrm{F}} \mathcal{Q}^0 = -J\sum\limits_{j,\alpha}(f_{j,\alpha}^{\dagger}f_{j+1,\alpha}+f_{j+1,\alpha}^{\dagger}f_{j,\alpha})\,.
\end{align}
Here, an allowed nearest-neighbor hop into an empty site preserves the order of the particles and hence their ordered spin sequence. With spin--charge separation, the projected hopping therefore acts only on the charge sector and can be expressed in terms of spinless-fermion operators $f_j$:
\begin{align}
\mathcal{Q}^0 H_{\mathrm{F}} \mathcal{Q}^0 = -J\sum\limits_{j}(f_{j}^{\dagger}f_{j+1}+f_{j+1}^{\dagger}f_{j})\,.
\label{eqn:A2}
\end{align}
The second-order term $\mathcal{Q}^0 H_{\mathrm{F}} \mathcal{Q}^1\dfrac{1}{\mathcal{Q}^1 H_{\mathrm{F}} \mathcal{Q}^1}\mathcal{Q}^1H_{\mathrm{F}} \mathcal{Q}^0$
{
\medmuskip=0mu
\thinmuskip=0mu
\small
\begin{align*}
= &\underbrace{\dfrac{J^2}{\mathcal{Q}^1 H_{\mathrm{F}}\mathcal{Q}^1}\sum\limits_{j,\alpha\beta}\mathcal{Q}^0\left( c_{j,\alpha}^{\dagger}e^{-i\theta n_{j}}c_{j+1,\alpha}\right)\mathcal{Q}^1\left( c_{j+1,\beta}^{\dagger}e^{+i\theta n_{j}} c_{j,\beta}\right)\mathcal{Q}^0}_{\crc{\text{a}}}
\\ + &\underbrace{\dfrac{J^2}{\mathcal{Q}^1 H_{\mathrm{F}}\mathcal{Q}^1}\sum\limits_{j,\alpha\beta}\mathcal{Q}^0\left(c_{j+1,\alpha}^{\dagger}e^{+i\theta n_{j}} c_{j,\alpha}\right)\mathcal{Q}^1\left( c_{j,\beta}^{\dagger}e^{-i\theta n_{j}}c_{j+1,\beta}\right)\mathcal{Q}^0}_{\crc{\text{b}}}
\\ + &\underbrace{\dfrac{J^2}{\mathcal{Q}^1 H_{\mathrm{F}}\mathcal{Q}^1}\sum\limits_{j,\alpha\beta}\mathcal{Q}^0\left( c_{j-1,\alpha}^{\dagger}e^{-i\theta n_{j-1}}c_{j,\alpha}\right)\mathcal{Q}^1\left( c_{j,\beta}^{\dagger}e^{-i\theta n_{j}}c_{j+1,\beta}\right)\mathcal{Q}^0}_{\crc{\text{c}}}
\\ + &\underbrace{\dfrac{J^2}{\mathcal{Q}^1 H_{\mathrm{F}}\mathcal{Q}^1}\sum\limits_{j,\alpha\beta}\mathcal{Q}^0\left( c_{j+2,\alpha}^{\dagger}e^{+i\theta n_{j+1}} c_{j+1,\alpha}\right)\mathcal{Q}^1\left( c_{j+1,\beta}^{\dagger}e^{+i\theta n_{j}} c_{j,\beta}\right)\mathcal{Q}^0}_{\crc{\text{d}}}
\end{align*}}
Only the tunneling term couples the two subspaces; to leading order, the Hamiltonian restricted to the high-energy subspace, $\mathcal Q^1H_{\mathrm F}\mathcal Q^1$, is approximated by its on-site interaction part. The energy denominator for each virtual process is thus the interaction energy of the intermediate doubly occupied site.

To simplify the expressions further, we define spin operators in the two-component $(\alpha,\beta)$ subspace as
\begin{align}
    \begin{split}
        & n_{j,(\alpha\beta)} = \bar{n}_{j,\beta}+\bar{n}_{j,\alpha}\,, \\
        & S_{j,(\alpha\beta)}^z = \bar{n}_{j,\beta}-\bar{n}_{j,\alpha}\,, \\
        & S_{j,(\alpha\beta)}^x = f_{j,\beta}^{\dagger}f_{j,\alpha}
        +f_{j,\alpha}^{\dagger} f_{j,\beta}\,, \\
        & S_{j,(\alpha\beta)}^y = -i \left(
            f_{j,\beta}^{\dagger}f_{j,\alpha}-f_{j,\alpha}^{\dagger}
            f_{j,\beta}
        \right)\,.
    \end{split}
    \label{eqn:A4}
\end{align}
Simplifying different parts of the second-order term:
{\small
 \medmuskip=0mu
\thinmuskip=0mu
\begin{flalign*}
\crc{\text{a}} =& \dfrac{J^2}{\mathcal{Q}^1 H_{\mathrm{F}}\mathcal{Q}^1}\sum\limits_{j,\alpha}\mathcal{Q}^0\left( c_{j,\alpha}^{\dagger}e^{-i\theta n_{j}}c_{j+1,\alpha}\right)\mathcal{Q}^1\left( c_{j+1,\alpha}^{\dagger}e^{+i\theta n_{j}} c_{j,\alpha}\right)\mathcal{Q}^0  
\\ +& \dfrac{J^2}{\mathcal{Q}^1 H_{\mathrm{F}}\mathcal{Q}^1}\smashoperator[r]{\sum\limits_{j,\alpha \neq \beta}}\mathcal{Q}^0\left( c_{j,\alpha}^{\dagger}e^{-i\theta n_{j}}c_{j+1,\alpha}\right)\mathcal{Q}^1\left( c_{j+1,\beta}^{\dagger}e^{+i\theta n_{j}} c_{j,\beta}\right)\mathcal{Q}^0
\\ =& \sum\limits_{j,\alpha \neq \beta}\dfrac{J^2}{U_{\alpha\beta}} \mathcal{Q}^0\left( c_{j,\alpha}^{\dagger}e^{-i\theta n_{j}}c_{j+1,\alpha}\right)n_{j+1,\beta}\left( c_{j+1,\alpha}^{\dagger}e^{+i\theta n_{j}} c_{j,\alpha}\right)\mathcal{Q}^0  
\\ +& \sum\limits_{j,\alpha \neq \beta}\dfrac{J^2}{U_{\alpha\beta}} \mathcal{Q}^0\left( c_{j,\alpha}^{\dagger}e^{-i\theta n_{j}}c_{j+1,\alpha}\right)\left( c_{j+1,\beta}^{\dagger}e^{+i\theta n_{j}} c_{j,\beta}\right)\mathcal{Q}^0
\\ =& \sum\limits_{j,\alpha \neq \beta} \dfrac{J^2}{U_{\alpha\beta}}\left(\bar{n}_{j,\alpha}\bar{n}_{j+1,\beta} - f_{j,\alpha}^{\dagger}f_{j,\beta}f_{j+1,\beta}^{\dagger}f_{j+1,\alpha}\right)\,,
\\ \crc{\text{b}} =& \sum\limits_{j,\alpha \neq \beta} \dfrac{J^2}{U_{\alpha\beta}}\left(\bar{n}_{j,\beta}\bar{n}_{j+1,\alpha} - f_{j,\beta}^{\dagger}f_{j,\alpha}f_{j+1,\alpha}^{\dagger}f_{j+1,\beta}\right)\,,\\
\crc{\text{a}} \;+&\; \crc{\text{b}} =\smashoperator[l]{\sum\limits_{j,\alpha < \beta}} \dfrac{2J^2}{U_{\alpha\beta}}\left(\bar{n}_j\bar{n}_{j+1}\right) \left(\bar{n}_{j,\alpha}\bar{n}_{j+1,\beta}+\bar{n}_{j,\beta}\bar{n}_{j+1,\alpha} \right.&\\
&\phantom{\qquad\qquad} \left. -f_{j,\alpha}^{\dagger}f_{j,\beta}f_{j+1,\beta}^{\dagger}f_{j+1,\alpha}
- f_{j,\beta}^{\dagger}f_{j,\alpha}f_{j+1,\alpha}^{\dagger}f_{j+1,\beta}\right)&\\
&\phantom{\;\;\,}=\smashoperator[l]{\sum\limits_{j,\alpha < \beta}} \dfrac{2J^2}{U_{\alpha\beta}}\left(\bar{n}_j\bar{n}_{j+1}\right)\left(\dfrac{n_{j,(\alpha\beta)}n_{j+1,(\alpha\beta)}-S^z_{j,(\alpha\beta)}S^z_{j+1,(\alpha\beta)}}{2} \right.&\\
&\left. \phantom{\qquad\qquad\qquad\quad} + \dfrac{-S^x_{j,(\alpha\beta)}S^x_{j+1,(\alpha\beta)}-S^y_{j,(\alpha\beta)}S^y_{j+1,(\alpha\beta)}}{2}\right)\,,\\
\crc{\text{a}} \;+&\; \crc{\text{b}} =J^2 \sum\limits_{j} (2\bar{n}_j\bar{n}_{j+1}) \left(\sum\limits_{\alpha<\beta}\dfrac{1}{U_{\alpha\beta}}(I-\mathcal{E}_{j,j+1})\hat{P}^{\alpha\beta}_{j,j+1}\right)\,,
\end{flalign*}
}where the exchange operator in the two-component subspace $\mathcal{E}_{j,j+1} = \big[n_{j,(\alpha\beta)}n_{j+1,(\alpha\beta)} +\sum\limits_{\kappa=x,y,z} S_{j,(\alpha\beta)}^{\kappa}S_{j+1,(\alpha\beta)}^{\kappa}\big]/2$.
{
\small
\medmuskip=0mu
\thinmuskip=0mu
\begin{flalign*}
\crc{\text{c}} = &\sum\limits_{j,\alpha \neq \beta}\dfrac{J^2}{U_{\alpha\beta}}\mathcal{Q}^0\left( c_{j-1,\alpha}^{\dagger}e^{-i\theta n_{j-1}}c_{j,\alpha}\right)n_{j,\beta}\left( c_{j,\alpha}^{\dagger}e^{-i\theta n_{j}}c_{j+1,\alpha}\right)\mathcal{Q}^0 
    \\ + &\sum\limits_{j,\alpha \neq \beta}\dfrac{J^2}{U_{\alpha\beta}}\mathcal{Q}^0\left( c_{j-1,\alpha}^{\dagger}e^{-i\theta n_{j-1}}c_{j,\alpha}\right)\left( c_{j,\beta}^{\dagger}e^{-i\theta n_{j}}c_{j+1,\beta}\right)\mathcal{Q}^0
    \\ = &\smashoperator{\sum\limits_{j,\alpha < \beta}}\;\Bigg[\dfrac{J^2e^{-i\theta}}{U_{\alpha\beta}}\left(f_{j-1,\alpha}^{\dagger}n_{j,\beta}f_{j+1,\alpha} - f_{j-1,\alpha}^{\dagger}f_{j,\beta}^{\dagger}f_{j,\alpha} f_{j+1,\beta}\right)
    \\& \phantom{\quad\;\,\,} +\dfrac{J^2e^{-i\theta}}{U_{\alpha\beta}}\left(f_{j-1,\beta}^{\dagger}n_{j,\alpha}f_{j+1,\beta} - f_{j-1,\beta}^{\dagger}f_{j,\alpha}^{\dagger}f_{j,\beta} f_{j+1,\alpha}\right)\Bigg]\,,\\
= &- J^2 \smashoperator{\sum\limits_{j}} \left( e^{-i\theta}f_{j-1}^{\dagger}n_{j}f_{j+1}\right) \left(\sum\limits_{\alpha<\beta}\dfrac{1}{U_{\alpha\beta}}(I-\mathcal{E}_{j,j+1})\hat{P}^{\alpha\beta}_{j,j+1}\right)\,,
\end{flalign*}}
In the last equality, to obtain a spin--charge separated form, we have employed
{
\small
\begin{align*}
\begin{split}
        f^{\dagger}_{j-1,\alpha}n_{j,\beta}&f_{j+1,\alpha} \\&= f^{\dagger}_{j-1,\alpha}\left(f^{\dagger}_{j,\beta}f_{j,\beta}\right)f_{j+1,\alpha}
        \\& = f^{\dagger}_{j-1,\alpha}f^{\dagger}_{j,\beta}\left(f_{j,\alpha}f^{\dagger}_{j,\alpha}\right)f_{j,\beta}\left(f_{j+1,\beta}f^{\dagger}_{j+1,\beta}\right)f_{j+1,\alpha}
        \\& = -\left(f^{\dagger}_{j-1,\alpha}f_{j,\alpha}\right)\left(f^{\dagger}_{j,\beta}f_{j+1,\beta}\right) f^{\dagger}_{j,\alpha}f_{j,\beta}f^{\dagger}_{j+1,\beta}f_{j+1,\alpha}
    \\&=-\left(f^{\dagger}_{j-1}f_{j}\right)\left(f^{\dagger}_{j}f_{j+1}\right) f^{\dagger}_{j,\alpha}f_{j,\beta}f^{\dagger}_{j+1,\beta}f_{j+1,\alpha}
    \\&=\left(f^{\dagger}_{j-1}f^{\dagger}_{j} f_{j}f_{j+1}\right)f^{\dagger}_{j,\alpha}f_{j,\beta}f^{\dagger}_{j+1,\beta}f_{j+1,\alpha}
    \\&=\left(f^{\dagger}_{j-1}n_jf_{j+1}\right)S_{j,(\alpha\beta)}^-S_{j+1,(\alpha\beta)}^+\,,
    \\f^{\dagger}_{j-1,\beta}n_{j,\alpha}&f_{j+1,\beta} =\left(f^{\dagger}_{j-1}n_jf_{j+1}\right)S_{j,(\alpha\beta)}^+S_{j+1,(\alpha\beta)}^-\,,
    \\-f^{\dagger}_{j-1,\alpha}f^{\dagger}_{j,\beta} &f_{j,\alpha}f_{j+1,\beta} \\&= -f^{\dagger}_{j-1,\alpha}f^{\dagger}_{j,\beta} f_{j,\alpha}\left(n_{j,\alpha}\right)f_{j+1,\beta}\left(n_{j+1,\beta} \right)
        \\& = -\left(-f^{\dagger}_{j-1,\alpha}f_{j,\alpha}\right) \left(f^{\dagger}_{j,\beta}f_{j+1,\beta} \right)n_{j,\alpha}n_{j+1,\beta} 
        \\& = \left(f^{\dagger}_{j-1}f_{j}\right) \left(f^{\dagger}_{j}f_{j+1} \right)n_{j,\alpha}n_{j+1,\beta} 
    \\& = -\left(f^{\dagger}_{j-1}n_{j}f_{j+1} \right)\dfrac{1}{4}\left(I-S_{j,(\alpha\beta)}^z\right)\left(I+S_{j+1,(\alpha\beta)}^z\right)\,,
    \\-f^{\dagger}_{j-1,\beta}f^{\dagger}_{j,\alpha} &f_{j,\beta}f_{j+1,\alpha} \\&= -\left(f^{\dagger}_{j-1}n_{j}f_{j+1} \right)\dfrac{1}{4}\left(I+S_{j,(\alpha\beta)}^z\right)\left(I-S_{j+1,(\alpha\beta)}^z\right)\,.
\end{split}
\end{align*}
}In the two-component $\alpha<\beta$ subspace, $S_{j,(\alpha\beta)}^-$ and $S_{j,(\alpha\beta)}^+$ are spin-lowering and spin-raising operators, respectively.
{
\small
\medmuskip=0mu
\thinmuskip=0mu
\begin{flalign*}
    \crc{\text{d}} = &- J^2 \smashoperator{\sum\limits_{j}}
    \left(
        e^{+i\theta}f_{j+2}^{\dagger}
        n_{j+1}f_{j}
    \right)
    \left(
        \sum\limits_{\alpha<\beta}\dfrac{1}{U_{\alpha\beta}}
        (I-\mathcal{E}_{j,j+1})\hat{P}^{\alpha\beta}_{j,j+1}
    \right)\,,\\
    \crc{\text{c}}\;+\;\crc{\text{d}} =& - J^2 \sum\limits_{j}
    \left(
        \text{\scriptsize$
        \begin{aligned}
            &e^{-i\theta}f_{j-1}^{\dagger}
             n_{j}f_{j+1}\\
            &+e^{+i\theta}f_{j+2}^{\dagger}
             n_{j+1}f_{j}
        \end{aligned}
        $}
    \right)
    \left(
        \sum\limits_{\alpha<\beta}\dfrac{1}{U_{\alpha\beta}}
        (I-\mathcal{E}_{j,j+1})\hat{P}^{\alpha\beta}_{j,j+1}
    \right)\,,&&
\end{flalign*}
}

Then, the second-order term in Eq.~\eqref{eqn:A1}, in a spin--charge-separated form is
{\medmuskip=1mu
 \thinmuskip=1mu
    \begin{align}
        =-J^2 \sum\limits_j
        \left(
            \text{\small$
                \begin{aligned}
                    2\bar{n}_j
                     \bar{n}_{j+1}
                    &- e^{-i\theta}f_{j-1}^{\dagger}
                       n_j
                       f_{j+1}
                    \\
                    &- e^{+i\theta}f_{j+2}^{\dagger}
                       n_{j+1}
                       f_j
                \end{aligned}
            $}
        \right)
        \left(
            \sum\limits_{\alpha<\beta}
            \dfrac{(I-\mathcal{E}_{j,j+1})}{U_{\alpha\beta}}
            \hat{P}^{\alpha\beta}_{j,j+1}
        \right)\,.
    \end{align}
}After the second-order expansion and spin--charge separation, the effective Hamiltonian becomes
{
\medmuskip=0mu
\thinmuskip=0mu
\begin{align}
H_{\mathrm{eff}} = -J\sum\limits_{j}\left(\text{\small$
                \begin{aligned}
                    &f_{j}^{\dagger}f_{j+1}\\
                    &+f_{j+1}^{\dagger}f_{j}
                \end{aligned}
            $}\right)-J^2 &\sum\limits_{j} \Bigg[\left(
            \text{\small$
                \begin{aligned}
                    2\bar{n}_j
                     \bar{n}_{j+1}
                    &- e^{-i\theta}f_{j-1}^{\dagger}
                       n_j
                       f_{j+1}
                    \\
                    &- e^{+i\theta}f_{j+2}^{\dagger}
                       n_{j+1}
                       f_j
                \end{aligned}
            $}
        \right)
\notag\\&\times \sum\limits_{\alpha<\beta}\dfrac{(I-\mathcal{E}_{j,j+1})}{U_{\alpha\beta}}\hat{P}^{\alpha\beta}_{j,j+1}\Bigg]\,.
\end{align}
}

We express the spin operators using the ordered-particle index $\ell$, distinguishing spin-chain from physical lattice sites:
\begin{align}
H_{\mathrm{eff}} = -&J\sum\limits_{j}\left(\text{\small$
                \begin{aligned}
                    &f_{j}^{\dagger}f_{j+1}\\
                    &+f_{j+1}^{\dagger}f_{j}
                \end{aligned}
            $}\right)
-\sum\limits_{\ell}\sum\limits_{\alpha<\beta}\dfrac{J^2}{U_{\alpha\beta}}  \Bigg\{\sum\limits_{j} \krondel \Bigg[\notag\\ &\left(
            \text{\footnotesize$
                \begin{aligned}
                    2\bar{n}_j
                     \bar{n}_{j+1}
                    &- e^{-i\theta}f_{j-1}^{\dagger}
                       n_j
                       f_{j+1}
                    \\
                    &- e^{+i\theta}f_{j+2}^{\dagger}
                       n_{j+1}
                       f_j
                \end{aligned}
            $}
        \right)
 (I-\mathcal{E}_{\ell,\ell+1})\hat{P}^{\alpha\beta}_{\ell,\ell+1}\Bigg]\Bigg\}\,.
\label{eqn:eff_FH}
\end{align}
Here, $j$ labels physical lattice sites, whereas $\ell$ labels particles in their spatial order and hence sites of the effective spin chain. The Kronecker projector $\delta_{{\textstyle\sum\nolimits_{i=1}^{j-1}}{\scriptstyle\overline{n}_i,\ell-1}}$ selects configurations with exactly $\ell-1$ particles to the left of site $j$. For an initially occupied pair $(j,j+1)$, this identifies the two particles with neighboring spins $(\ell,\ell+1)$ in the effective spin chain.

We take the spin-independent tunneling term in
Eq.~\eqref{eqn:eff_FH} as the unperturbed charge Hamiltonian,
\begin{align}
    H_{\mathrm{C}} = -J\sum\limits_{j}(f_{j}^{\dagger}f_{j+1}+f_{j+1}^{\dagger}f_{j})\,,
\end{align}
and treat terms of order $J^2/U_{\alpha\beta}$ perturbatively.
Let $\ket{\varphi}$ denote the charge ground state with energy $E_0$. Since $H_{\mathrm C}$ leaves the spin wave function $\ket{\chi}$ undetermined, its ground-state manifold is spin degenerate. Degenerate perturbation theory gives the spin-chain Hamiltonian for anyons ($\expval{H_{\mathrm{eff}}}{\varphi}$): 
{
\medmuskip=0mu
\thinmuskip=0mu
\begin{align}
H_{\mathrm{SC:F}}  = E_0- J^2\sum\limits_{\ell}\mathcal{C}_\ell(\theta) \left(\sum\limits_{\alpha<\beta}\dfrac{1}{U_{\alpha\beta}}(I-\mathcal{E}_{\ell,\ell+1})\hat{P}^{\alpha\beta}_{\ell,\ell+1}\right)\,,
\end{align}
}where the bond coefficients, 
\begin{align*}
    \mathcal{C}_\ell(\theta) = \expval{\sum\limits_{j} \krondel \left(\begin{aligned}2\bar{n}_j\bar{n}_{j+1}&-e^{+i\theta}f_{j+2}^{\dagger}\bar{n}_{j+1}f_{j}
    \\&-e^{-i\theta}f_{j-1}^{\dagger}\bar{n}_{j}f_{j+1}\end{aligned}\right)}{\varphi}\,,
\end{align*}
of the operator acting on the spin wave function, contains all the charge degrees of freedom. This completes the GESC description of the $N$-component fermion-type Anyon--Hubbard model.

\begin{widetext}
\section{\label{app:bosonic_derivation}GESC Hamiltonian for boson-type anyons}
As in the fermion-type case, we derive the boson-type GESC Hamiltonian from the mapped Bose--Hubbard Hamiltonian in Eq.~\eqref{eqn:boseH}. Within the low-energy subspace, a Jordan--Wigner transformation expresses the $N$-component bosons in terms of projected fermions,
\begin{align}
    b_{j,\alpha} = \exp(i\pi\sum_{i<j}n_i) \exp(i\pi\sum_{\gamma<\alpha}n_{j,\gamma}) f_{j,\alpha}\,,
\end{align}
where the combination enforces at most one particle per site.

The zeroth-order term [Eq.~\eqref{eqn:A1}] is identical to that for fermion-type anyons:
\begin{align}
    \mathcal{Q}^0 H_{\mathrm{B}}\mathcal{Q}^0 =  -J\sum\limits_{j}\left( f_{j}^{\dagger}f_{j+1}+  f_{j+1}^{\dagger}f_{j}\right)\,.
\end{align}

The second-order term $\mathcal{Q}^0 H_{\mathrm{B}}\mathcal{Q}^1\dfrac{1}{\mathcal{Q}^1 H_{\mathrm{B}}\mathcal{Q}^1}\mathcal{Q}^1 H_{\mathrm{B}}\mathcal{Q}^0$
{\small
\begin{align*}
    = &\underbrace{\dfrac{J^2}{\mathcal{Q}^1 H_{\mathrm{B}}\mathcal{Q}^1}\sum\limits_{j,\alpha\beta} \mathcal{Q}^0\left( b_{j,\alpha}^{\dagger}b_{j+1,\alpha} e^{-i\theta n_{j}}\right) \mathcal{Q}^1\left( e^{+i\theta n_{j}} b_{j+1,\beta}^{\dagger}b_{j,\beta}\right) \mathcal{Q}^0}_{\crc{\text{a}}}
     +\underbrace{\dfrac{J^2}{\mathcal{Q}^1 H_{\mathrm{B}}\mathcal{Q}^1}\sum\limits_{j,\alpha\beta} \mathcal{Q}^0\left(e^{+i\theta n_{j}} b_{j+1,\alpha}^{\dagger}b_{j,\alpha}\right) \mathcal{Q}^1\left( b_{j,\beta}^{\dagger}b_{j+1,\beta} e^{-i\theta n_{j}}\right) \mathcal{Q}^0}_{\crc{\text{b}}}
     \\  + &\underbrace{\dfrac{J^2}{\mathcal{Q}^1 H_{\mathrm{B}}\mathcal{Q}^1}\sum\limits_{j,\alpha\beta} \mathcal{Q}^0\left( b_{j-1,\alpha}^{\dagger}b_{j,\alpha} e^{-i\theta n_{j-1}}\right) \mathcal{Q}^1\left( b_{j,\beta}^{\dagger}b_{j+1,\beta} e^{-i\theta n_{j}}\right) \mathcal{Q}^0}_{\crc{\text{c}}}
     + \underbrace{\dfrac{J^2}{\mathcal{Q}^1 H_{\mathrm{B}}\mathcal{Q}^1}\sum\limits_{j,\alpha\beta} \mathcal{Q}^0\left( e^{+i\theta n_{j+1}} b_{j+2,\alpha}^{\dagger}b_{j+1,\alpha}\right) \mathcal{Q}^1\left( e^{+i\theta n_{j}} b_{j+1,\beta}^{\dagger}b_{j,\beta}\right) \mathcal{Q}^0}_{\crc{\text{d}}}
\end{align*}
}Unlike fermions, bosons allow same-component multiple occupancy in the high-energy subspace. Thus, we next resolve the intermediate-state propagator by explicitly identifying the virtual high-energy configurations selected by $\mathcal{Q}^1$. The corresponding energy denominator depends on the nature of the intermediate doubly occupied state: $V_{\alpha}$ for same-component pairs and $U_{\alpha\beta}$ for different-component pairs. Accordingly,
\begin{align*}
    \sum_{\alpha,\beta} b_{j+1,\alpha}\dfrac{\mathcal{Q}^1}{\mathcal{Q}^1H_{\mathrm{B}}\mathcal{Q}^1}b_{j+1,\beta}^{\dagger} = \sum_{\alpha} \dfrac{b_{j+1,\alpha} (n_{j+1,\alpha}-1) b_{j+1,\alpha}^{\dagger}}{V_{\alpha}} + \sum_{\alpha\neq \beta} \dfrac{b_{j+1,\alpha} n_{j+1,\beta} b_{j+1,\alpha}^{\dagger}}{U_{\alpha\beta}} + \sum_{\alpha\neq \beta} \dfrac{b_{j+1,\alpha} b_{j+1,\beta}^{\dagger}}{U_{\alpha\beta}}\,.
\end{align*}

The two-site contributions then simplify to
{\medmuskip=0mu
\thinmuskip=0mu
\begin{flalign*}
    \crc{\text{a}}  = &\sum_{j,\alpha }J^2 \mathcal{Q}^0\left[ \dfrac{1}{V_{\alpha}}2n_{j,\alpha}n_{j+1,\alpha} 
        +\sum_{\beta \neq \alpha}\dfrac{1}{U_{\alpha\beta}} \left(n_{j,\alpha}n_{j+1,\beta}+b_{j,\alpha}^{\dagger} b_{j,\beta} b_{j+1,\beta}^{\dagger} b_{j+1,\alpha} \right)\right]\mathcal{Q}^0\,,&&
    \\\crc{\text{b}}  = &\sum_{j,\alpha }J^2 \mathcal{Q}^0\left[\dfrac{1}{V_{\alpha}}2n_{j,\alpha}n_{j+1,\alpha} 
        +\sum_{\beta \neq \alpha}\dfrac{1}{U_{\alpha\beta}} \left(n_{j,\beta}n_{j+1,\alpha}+b_{j,\beta}^{\dagger} b_{j,\alpha} b_{j+1,\alpha}^{\dagger} b_{j+1,\beta} \right)\right]\mathcal{Q}^0\,,&&    
\end{flalign*}
\begin{flalign*}
\crc{\text{a}}\;+\;\crc{\text{b}}  = &\sum_{j,\alpha<\beta }\dfrac{2J^2}{U_{\alpha\beta}} \mathcal{Q}^0\left[\begin{aligned}&2\mu_{\alpha;\beta} n_{j,\alpha}n_{j+1,\alpha}  + 2\mu_{\beta;\alpha} n_{j,\beta}n_{j+1,\beta} \\&+n_{j,\alpha}n_{j+1,\beta} + n_{j,\beta}n_{j+1,\alpha}\\&+b_{j,\alpha}^{\dagger} b_{j,\beta} b_{j+1,\beta}^{\dagger} b_{j+1,\alpha} + b_{j,\beta}^{\dagger} b_{j,\alpha} b_{j+1,\alpha}^{\dagger} b_{j+1,\beta}\end{aligned}\right]\mathcal{Q}^0
    \\ = &\sum\limits_{j,\alpha<\beta}J^2\dfrac{(2\bar{n}_j\bar{n}_{j+1})}{U_{\alpha\beta}}\left\{I + \mathcal{E}_{j,j+1} + \left(\dfrac{\mu_{\alpha;\beta}+\mu_{\beta;\alpha}}{2}-1\right)\left(I+S_{j,(\alpha\beta)}^zS_{j+1,(\alpha\beta)}^z\right) + \dfrac{\mu_{\beta;\alpha}-\mu_{\alpha;\beta}}{2}\left(S_{j,(\alpha\beta)}^z+S_{j+1,(\alpha\beta)}^z\right)\right\}\hat{P}_{j,j+1}^{\alpha\beta}\,,&&
\end{flalign*}}
where $\mu_{\alpha;\beta} = {U_{\alpha\beta}}/{[(N-1)V_{\alpha}]}$ and $\mu_{\beta;\alpha} = {U_{\alpha\beta}}/{[(N-1)V_{\beta}]}$.\\

The three-site contributions are
\begin{flalign*}
    \crc{\text{c}}  = &\sum_{j,\alpha <\beta }\dfrac{J^2}{U_{\alpha\beta}}e^{-i\theta } \mathcal{Q}^0\left[\begin{aligned}& 
    2\mu_{\alpha;\beta}b_{j-1,\alpha}^{\dagger}n_{j,\alpha} b_{j+1,\alpha} + 2\mu_{\beta;\alpha}b_{j-1,\beta}^{\dagger}n_{j,\beta} b_{j+1,\beta}
     \\&+ b_{j-1,\alpha}^{\dagger}n_{j,\beta} b_{j+1,\alpha} + b_{j-1,\beta}^{\dagger}n_{j,\alpha} b_{j+1,\beta}\\&+b_{j-1,\alpha}^{\dagger}b_{j,\beta}^{\dagger} b_{j,\alpha} b_{j+1,\beta} +b_{j-1,\beta}^{\dagger}b_{j,\alpha}^{\dagger} b_{j,\beta} b_{j+1,\alpha}\end{aligned}
    \right]\mathcal{Q}^0\,,
    \\\crc{\text{d}}  = &\sum_{j,\alpha <\beta }\dfrac{J^2}{U_{\alpha\beta}}e^{+i\theta } \mathcal{Q}^0\left[\begin{aligned}& 
    2\mu_{\alpha;\beta}b_{j+2,\alpha}^{\dagger}n_{j+1,\alpha} b_{j,\alpha} + 2\mu_{\beta;\alpha}b_{j+2,\beta}^{\dagger}n_{j+1,\beta} b_{j,\beta}
     \\&+ b_{j+2,\alpha}^{\dagger}n_{j+1,\beta} b_{j,\alpha} + b_{j+2,\beta}^{\dagger}n_{j+1,\alpha} b_{j,\beta}\\&+b_{j+2,\alpha}^{\dagger}b_{j+1,\beta}^{\dagger} b_{j+1,\alpha} b_{j,\beta} +b_{j+2,\beta}^{\dagger}b_{j+1,\alpha}^{\dagger} b_{j+1,\beta} b_{j,\alpha}\end{aligned}
    \right]\mathcal{Q}^0\,,&&
\end{flalign*}
As in the fermion-type case, we perform spin--charge separation of the three-site terms. For $\alpha<\beta$, representative identities are
\begin{align*}
    \mathcal{Q}^0\left[b_{j-1,\alpha}^{\dagger}n_{j,\alpha} b_{j+1,\alpha} \right]\mathcal{Q}^0 &  = \mathcal{Q}^0\left[\left(b_{j-1}^{\dagger}n_{j} b_{j+1}\right) n_{j,\alpha}n_{j+1,\alpha}\right]\mathcal{Q}^0 = -\left[\left(f_{j-1}^{\dagger}\bar{n}_{j} f_{j+1}\right) n_{j,\alpha}n_{j+1,\alpha}\right]\,,
    \\\mathcal{Q}^0\left[b_{j-1,\alpha}^{\dagger}n_{j,\beta} b_{j+1,\alpha} \right]\mathcal{Q}^0 &  = -\left[\left(f_{j-1}^{\dagger}\bar{n}_{j} f_{j+1}\right) S^{-}_{j,(\alpha\beta)}S^{+}_{j+1,(\alpha\beta)}\right]\,,
    \\\mathcal{Q}^0\left[b_{j-1,\alpha}^{\dagger}b_{j,\beta}^{\dagger} b_{j,\alpha} b_{j+1,\beta}\right]\mathcal{Q}^0 &  = -\left[\left(f_{j-1}^{\dagger}\bar{n}_{j} f_{j+1}\right) n_{j,\alpha}n_{j+1,\beta}\right]\,.
\end{align*}
The remaining terms follow analogously, together with their Hermitian-conjugate counterparts in $\crc{\text{d}}$.\\

Combining the spin--charge-separated three-site contributions, we obtain
\begin{flalign*}
\crc{\text{c}} + \crc{\text{d}} = &
-\sum\limits_{j,\alpha<\beta}\dfrac{J^2}{U_{\alpha\beta}}(e^{-i\theta}f_{j-1}^{\dagger}\bar{n}_{j}f_{j+1}+ e^{+i\theta}f_{j+2}^{\dagger}\bar{n}_{j+1}f_{j})\left[\begin{aligned}I + \mathcal{E}_{j,j+1} &+ \left(\dfrac{\mu_{\alpha;\beta}+\mu_{\beta;\alpha}}{2}-1\right)\left(I+S_{j,(\alpha\beta)}^zS_{j+1,(\alpha\beta)}^z\right) \\&+ \dfrac{\mu_{\beta;\alpha}-\mu_{\alpha;\beta}}{2}\left(S_{j,(\alpha\beta)}^z+S_{j+1,(\alpha\beta)}^z\right)\end{aligned}\right]\hat{P}_{j,j+1}^{\alpha\beta}\,.&&
\end{flalign*}

Combining the two- and three-site contributions gives the second-order term in Eq.~\eqref{eqn:A1}, 
\begin{align}
    \sum\limits_{j,\alpha<\beta}\dfrac{J^2}{U_{\alpha\beta}}\left(\begin{aligned}2\bar{n}_j\bar{n}_{j+1} &-e^{-i\theta}f_{j-1}^{\dagger}\bar{n}_{j}f_{j+1}\\&- e^{+i\theta}f_{j+2}^{\dagger}\bar{n}_{j+1}f_{j}\end{aligned}\right)\left[\begin{aligned}I + \mathcal{E}_{j,j+1} &+ \left(\dfrac{\mu_{\alpha;\beta}+\mu_{\beta;\alpha}}{2}-1\right)\left(I+S_{j,(\alpha\beta)}^zS_{j+1,(\alpha\beta)}^z\right) \\&+ \dfrac{\mu_{\beta;\alpha}-\mu_{\alpha;\beta}}{2}\left(S_{j,(\alpha\beta)}^z+S_{j+1,(\alpha\beta)}^z\right)\end{aligned}\right]\hat{P}_{j,j+1}^{\alpha\beta}\,.
\end{align}
Proceeding as in the fermion-type case, we obtain the effective spin-chain Hamiltonian for $N$-component boson-type anyons as
{\medmuskip=0mu
\thinmuskip=2mu
\begin{align}
H_{\mathrm{SC:B}} =  E_0- J^2\sum\limits_{\ell}\mathcal{C}_{\ell}(\theta)\left[\sum\limits_{\alpha<\beta}\dfrac{1}{U_{\alpha\beta}}\left[\left(I+\mathcal{E}_{\ell,\ell+1}\right) + \left(\tfrac{\mu_{\alpha;\beta}+\mu_{\beta;\alpha}}{2}-1\right)\left(I+S_{\ell,\left(\alpha\beta\right)}^zS_{\ell+1,\left(\alpha\beta\right)}^z\right) + \tfrac{\mu_{\beta;\alpha}-\mu_{\alpha;\beta}}{2}\left(S_{\ell,\left(\alpha\beta\right)}^z+S_{\ell+1,\left(\alpha\beta\right)}^z\right) \right] \hat{P}^{\alpha\beta}_{\ell,\ell+1}\right]\;, 
\end{align}
}where the coefficient $\mathcal{C}_{\ell}(\theta)$, defined in Eq.~\eqref{eqn:HSCF1}, is identical to that for fermion-type anyons.
\clearpage
\end{widetext}

\begin{figure*}
\centering
\includegraphics{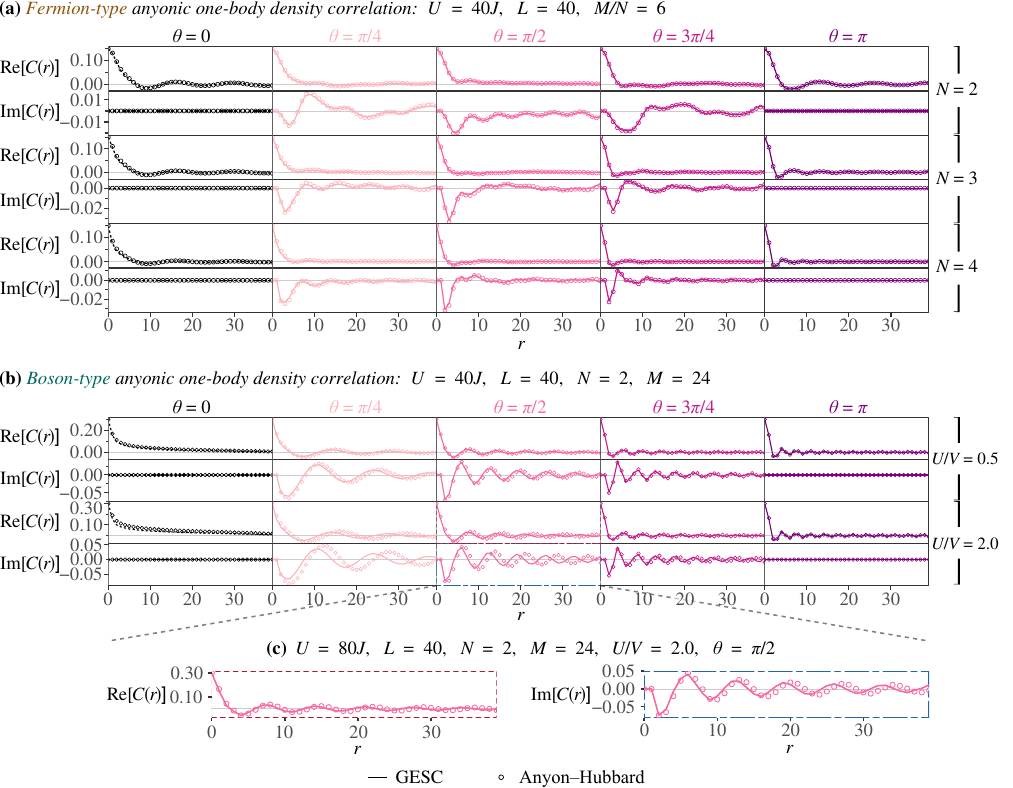}
\caption{\label{fig:gesc_benchmark_Cr}\textbf{GESC captures statistics-dependent oscillations and the decay of one-body density correlations.} Real and imaginary parts of $C(r)$ from GESC (curves) and the full Anyon--Hubbard model (circles). Panel (c) doubles both interactions relative to the highlighted case in (b), improving agreement at fixed $U/V=2$ and unchanged particle numbers. Separation $r$ is measured in lattice sites.}
\end{figure*}

\begin{figure*}
\centering
\includegraphics{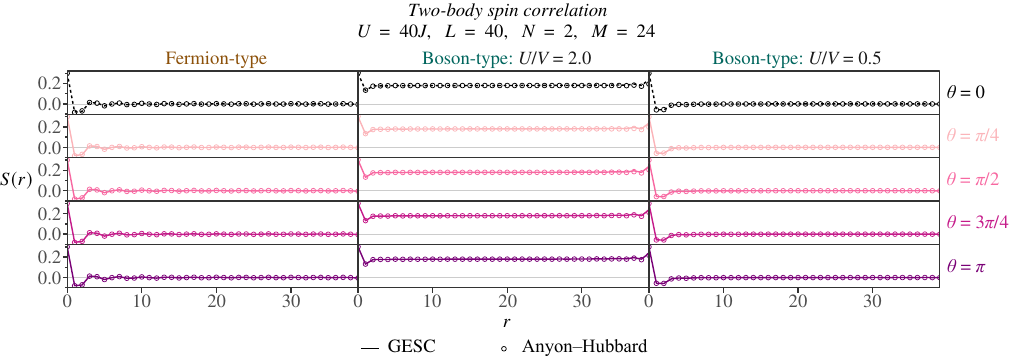}
\caption{\label{fig:gesc_benchmark_spin_Sr}\textbf{Interaction anisotropy sets the character of spin correlations across exchange regimes.} Spatially averaged diagonal spin correlations $S(r)$, comparing GESC (curves) with the full Anyon--Hubbard model (open circles). GESC captures the weak variation with statistical phase and the contrast between negative short-range correlations and the extended positive correlations at $U/V=2$. Separation $r$ is measured in lattice sites.}
\end{figure*}

\section{\label{app:gesc_realspace_benchmarks}Real-space benchmarks of the GESC formalism}
We benchmark GESC against the full Anyon--Hubbard (AH) model using the spatially averaged one-body density correlations $C(r)= (L-r)^{-1}\sum_{m} C(m,m+r)$ and spin correlations $S(r) = (L-r)^{-1}\sum_{m} S(m,m+r)$ 
as functions of the site separation $r$. These comparisons test the real-space correlations underlying the momentum distribution and spin structure factor. All benchmarks use $L=40$ sites with open boundary conditions. For boson-type anyons, $U$ and $V$ denote the on-site intercomponent and intracomponent interactions, respectively. For these benchmarks, no constraint is imposed on the spin-component populations, so that both the GESC and the full Anyon--Hubbard calculations determine their ground states within the same unrestricted Hilbert space.

For fermion-type anyons, \hyperref[fig:gesc_benchmark_Cr]{Fig.~\ref{fig:gesc_benchmark_Cr}(a)} compares the two approaches at $U=40J$ for $N=2$, $3$, and $4$, with $M=6N$. The GESC and AH results are essentially identical for both the real and imaginary parts of $C(r)$. GESC reproduces the characteristic oscillations and decay of $\Re[C(r)]$, which determine the position and sharpness of the corresponding nonanalytic momentum-space features. It also captures the finite $\Im[C(r)]$ at intermediate $\theta$, which generates the asymmetry in the momentum distribution. The imaginary part vanishes at $\theta=0$ and $\pi$, consistent with the restoration of momentum symmetry at these endpoints. The corresponding Anyon--Hubbard ground-state physics and its momentum-space signatures are discussed in Ref.~\cite{basak_anyon1} (Secs.~III~C--D, and Figs.~2--5 therein).

The boson-type benchmarks in \hyperref[fig:gesc_benchmark_Cr]{Fig.~\ref{fig:gesc_benchmark_Cr}(b)} use $N=2$, $M=24$, and $U=40J$, with $V=20J$ and $80J$. The benchmarks being restricted to two components reflects the substantially larger computational cost of the full Bose--Hubbard calculation: even when the local Hilbert space is truncated to occupancies of at most two particles per site, $N=2$ already requires six local basis states, with the dimension increasing quadratically with $N$ as $d_{\mathrm{loc}}=(N+1)(N+2)/2$. GESC reproduces the nonoscillatory power-law decay of $\Re[C(r)]$ at $\theta=0$, the emergence of decaying oscillations with increasing $\theta$, and the finite $\Im[C(r)]$ at intermediate phases. The latter vanishes at $\theta=0$ and $\pi$. Differences in oscillation amplitudes and extrema remain at intermediate phases, particularly in the imaginary part for $V=20J$. Agreement improves for $V=80J$, consistent with stronger suppression of same-component double occupancy. Doubling both interactions to $U=80J$ and $V=40J$ at fixed $U/V=2$ substantially improves agreement at $\theta=\pi/2$ \hyperref[fig:gesc_benchmark_Cr]{Figure~\ref{fig:gesc_benchmark_Cr}(c)}. GESC closely reproduces the short-distance extrema of $C(r)$, with only small oscillation phase and amplitude differences remaining at larger separations.

The spin-correlation benchmarks at $U=40J$ are shown in \hyperref[fig:gesc_benchmark_spin_Sr]{Fig.~\ref{fig:gesc_benchmark_spin_Sr}} [same parameters as \hyperref[fig:gesc_benchmark_Cr]{Fig.~\ref{fig:gesc_benchmark_Cr}(b)}]. For fermion-type anyons, $S(r)$ exhibits damped oscillations about zero for all $\theta$, characteristic of antiferromagnetic ordering. For boson-type anyons with $U/V=0.5$, a similar alternating-sign structure is recovered, although the oscillations are less pronounced; increasing $V$ suppresses same-component double occupancy and drives the system toward the hard-core limit, where it mimics fermionic spin correlations. In contrast, for $U/V=2$ the correlations remain positive and nonoscillatory over the displayed range, signaling ferromagnetic spin correlations. GESC and AH results coincide across different spin-correlation regimes and for all statistical phases shown. $S(r)$ varies weakly with $\theta$, unlike $C(r)$. Without an explicit statistical string, $S(r)$ depends on statistics through the spin state and remains less sensitive than $C(r)$ to residual GESC--AH differences.

Together, the benchmarks show GESC capturing dominant real-space correlations in the strong-interaction regime. Residual differences, visible in boson-type one-body density correlations, decrease as interaction strengths increase, consistent with finite-interaction corrections beyond the mapping.

\begin{figure*}
\centering
\includegraphics{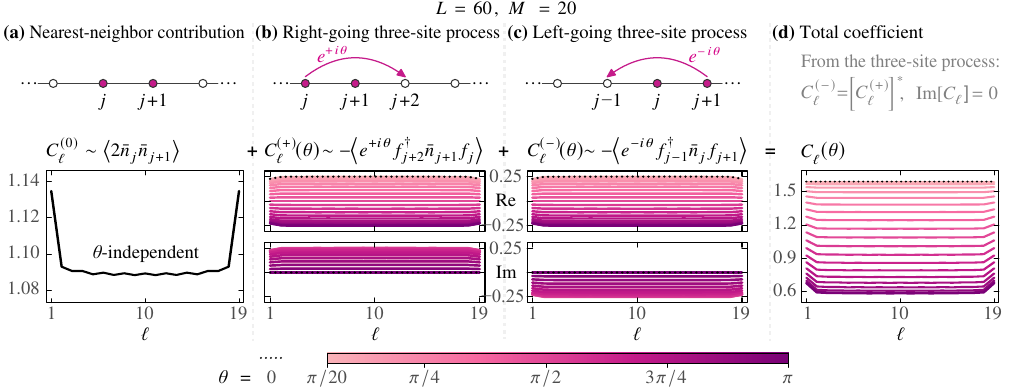}
\caption{\label{fig:Cl_decomposition}\textbf{Microscopic contributions to the bond-dependent spin-exchange coefficients $\bm{\mathcal{C}_\ell(\theta)}$.}  Schematics and numerical decomposition into (a) the nearest-neighbor contribution, (b,c) the real (upper) and imaginary (lower) parts of the right- and left-going three-site contributions, and (d) the total coefficient. The nearest-neighbor contribution is independent of $\theta$, while the three-site processes carry statistical phases $e^{\pm i\theta}$. Their imaginary parts cancel, leaving a real, $\theta$-dependent total coefficient. Filled circles for occupied and and open circles for empty sites.}
\end{figure*}

\begin{figure}
\centering
\includegraphics{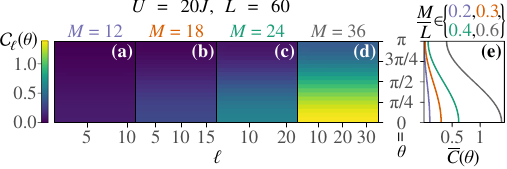}
\caption{\label{fig:spinexchange_clcoeff}\textbf{Fractional exchange statistics rescales, not reshapes spin exchange.} Dimensionless ground-state GESC coefficients $\mathcal C_\ell(\theta)$ and (last column) the mean
$\overline{\mathcal C}(\theta)=(M-1)^{-1}
\sum_{\ell=1}^{M-1}\mathcal C_\ell(\theta)$. The coefficients $\mathcal C_\ell(\theta)$ approximately retain their bond dependence while changing in magnitude. This rescaling is consistent with the weak statistical dependence of spin-ground-state correlations.}
\end{figure}

\section{\label{app:gesc_statemediated_analysis}State-mediated statistical dependence}

Here, we quantify the state-mediated dependence on the statistical phase within the GESC description. The charge Hamiltonian and charge wave function $\varphi$ are $\theta$ independent for both fermion- and boson-type anyons, so the statistical dependence of the spin state enters through the bond-dependent spin-exchange coefficients $\mathcal C_\ell(\theta)$. We focus on whether this dependence changes the spatial structure of the $\mathcal C_\ell(\theta)$ profile sufficiently to reconstruct the spin state.

Define operators
\begin{align}
    \begin{split}
        \hat{\mathcal{C}}^{(0)}_{\ell} &= \textstyle\sum\limits_{j} \krondelwide 2\bar{n}_j\bar{n}_{j+1}\,,
        \\\hat{\mathcal{C}}^{(+)}_{\ell} &= \textstyle\sum\limits_{j} \krondelwide f_{j+2}^{\dagger}\bar{n}_{j+1}f_{j}\,,
        \\\hat{\mathcal{C}}^{(-)}_{\ell} &= \textstyle\sum\limits_{j} \krondelwide f_{j-1}^{\dagger}\bar{n}_{j}f_{j+1}\,.
    \end{split}
\end{align}
These operators can be used to decompose the $\theta$-dependent spin-exchange coefficients into 
\begin{align}
\begin{split}
    \mathcal{C}^{(0)}_{\ell} &=  \expval{\hat{\mathcal{C}}^{(0)}_{\ell}}{\varphi}\,,\\
    \mathcal{C}^{(+)}_{\ell}(\theta) &= -\expval{e^{+i\theta}\hat{\mathcal{C}}^{(+)}_{\ell}}{\varphi}\,,\\
    \mathcal{C}^{(-)}_{\ell}(\theta) &= -\expval{e^{-i\theta}\hat{\mathcal{C}}^{(+)}_{\ell}}{\varphi}\,,
    \label{eqn:CLdecop}   
\end{split}
\end{align}
where $\mathcal C_\ell(\theta) = \mathcal{C}^{(0)}_{\ell}  + \mathcal{C}^{(+)}_{\ell}(\theta) + \mathcal{C}^{(-)}_{\ell}(\theta)$. Generalizing this to time-dependent spin-exchange coefficients to incorporate dynamics 
\begin{align}
    \begin{split}
        \mathcal{C}^{(0)}_{\ell}(t) &=  \expval{\hat{\mathcal{C}}^{(0)}_{\ell}}{\varphi(t)}\,,\\
        \mathcal{C}^{(\pm)}_{\ell}(\theta,t) &= -\expval{e^{\pm i\theta}\hat{\mathcal{C}}^{(\pm)}_{\ell}}{\varphi(t)}\,,\\
        \mathcal C_\ell(\theta,t) &= \mathcal{C}^{(0)}_{\ell}(t)  + \mathcal{C}^{(+)}_{\ell}(\theta,t) + \mathcal{C}^{(-)}_{\ell}(\theta,t)\,.
    \end{split}
    \label{eqn:CLdecop_t} 
\end{align}
Hereafter, we use Eq.~\eqref{eqn:CLdecop_t} for both the ground-state and dynamical analyses. For the ground-state analysis, $\ket{\varphi(0)}$ is the ground state of the corresponding charge Hamiltonian; we evaluate the coefficients at $t=0$ and drop the time variable, writing $\mathcal C_\ell^{(0)}$ and $\mathcal C_\ell^{(\pm)}(\theta)$. For the dynamical analysis, $\ket{\varphi(0)}$ denotes the prepared initial charge state, and we retain the time variable: $\mathcal C_\ell^{(0)}(0)$ and $\mathcal C_\ell^{(\pm)}(\theta,0)$ denote the initial coefficients, while $\mathcal C_\ell^{(0)}(t)$ and $\mathcal C_\ell^{(\pm)}(\theta,t)$ describe their subsequent evolution.

In the following, to separate the reflection symmetric part from the asymmetric part that can generate a right--left asymmetry, we express the $\theta$-dependent terms in $\mathcal C_{\ell}(\theta,t)$ as
\begin{align}
    \mathcal C_\ell^{(\pm)}(\theta,t)
    =
    e^{\pm i\theta}\mathcal C_\ell^{(\pm)}(0,t)\,.
    \label{eqn:thetato0}
\end{align}
Since
$\mathcal C_\ell^{(-)}(0,t)
=
[\mathcal C_\ell^{(+)}(0,t)]^*$,
we obtain
\begin{align}
\begin{split}
\mathcal C_\ell(\theta,t)
=
\mathcal C_\ell^{(0)}(t)
&+
2\Re[\mathcal C_\ell^{(+)}(0,t)]\cos\theta\\
&-
2\Im[\mathcal C_\ell^{(+)}(0,t)]\sin\theta\,.
\label{eq:Ctheta}
\end{split}
\end{align}
Let $\mathcal R$ denote reflection about the center of the lattice. With $j'=L-j$, reflection maps the neighboring sites $(j,j+1)$ to $(j'+1,j')$. We assume that the charge state $\ket{\varphi(0)}$, whether a ground state or an initial state for time evolution, is invariant under reflection up to a global phase $e^{i\alpha}$, and that the charge Hamiltonian $H_{\mathrm C}$ is reflection symmetric:
\begin{align}
\mathcal R\ket{\varphi(0)}
=
e^{i\alpha}\ket{\varphi(0)},
\qquad
\comm{\mathcal R}{H_{\mathrm C}}=0\,.
\end{align}
These conditions ensure that the charge state remains reflection symmetric throughout the evolution. At each time, the expectation value of any operator therefore equals that of its reflected counterpart,
\begin{align}
\begin{split}
\mathcal R\ket{\varphi(t)}
&=
e^{i\alpha}\ket{\varphi(t)},\\
\expval{\mathcal R\hat O\mathcal R^\dagger}{\varphi(t)}
&=
\expval{\hat O}{\varphi(t)}.
\end{split}
\label{eqn:operator_symmetric}
\end{align}
For the remainder of this analysis, we consider ground states or initial charge states and charge Hamiltonians that satisfy these reflection-symmetry conditions. Consider a setup with $\ell-1$ particles to the left of site $j$. Under reflection, these particles are then, mapped to the right of site $j'+1$. In addition, each of the three contributions $\mathcal C^{(0)}_\ell(t), \mathcal C^{(\pm)}_\ell(\theta,t)$ is non-zero only when sites $j$ and $j+1$ are both occupied. Reflection maps this occupied pair to sites $j'$ and $j'+1$. Since the reflected configuration has $\ell-1$ particles to the right of this pair and two particles within it, the remaining $M-(\ell-1)-2=M-\ell-1$ particles lie to the left of site $j'$. Hence, the Kronecker delta function that counts the particles in $\mathcal C_\ell(\theta,t)$ transforms as
\begin{align}
\mathcal R \krondel
\mathcal R^\dagger
=
\delta_{\smashoperator{{\textstyle\sum}\limits_{k=j'+2}^{L}}{\scriptstyle\overline{n}_k,\ell-1}}
=
\delta_{\smashoperator[r]{{\textstyle\sum}\limits_{k=1}^{j'-1}}{\scriptstyle\overline{n}_k,M-\ell-1}}.
\end{align}
Thus the reflected process acts on spin-chain bond $M-\ell$. Before obtaining the reflected coefficients, we first map the corresponding operators:
\begin{align}
    \begin{aligned}
        \mathcal R \hat{\mathcal{C}}^{(0)}_{\ell} \mathcal R^{\dagger} &= \mathcal R \left[\textstyle\sum\limits_{j} \krondelwide 2\bar{n}_j\bar{n}_{j+1}\right]\mathcal R^{\dagger} 
        \\&= \textstyle\sum\limits_{j'}\delta_{{\textstyle\sum\nolimits_{k=1}^{j'-1}}{\scriptstyle\overline{n}_k,M-\ell-1}} 2\bar{n}_{j'}\bar{n}_{j'+1}
        =  \hat{\mathcal{C}}^{(0)}_{M-\ell}\,,
    \end{aligned}
    \label{eqn:mapCl0}\raisetag{36pt}\\
    \begin{aligned}
        \mathcal R \hat{\mathcal{C}}^{(+)}_{\ell} \mathcal R^{\dagger} &= \mathcal R \left[\textstyle\sum\limits_{j} \krondelwide f_{j+2}^{\dagger}\bar{n}_{j+1}f_{j}\right]\mathcal R^{\dagger} 
        \\&=\textstyle\sum\limits_{j'}\delta_{{\textstyle\sum\nolimits_{k=1}^{j'-1}}{\scriptstyle\overline{n}_k,M-\ell-1}} f_{j'-1}^{\dagger}\bar{n}_{j'}f_{j'+1}
        = \hat{\mathcal{C}}^{(-)}_{M-\ell}\,,
    \end{aligned}
    \label{eqn:mapClp}\raisetag{36pt}\\
    \begin{aligned}
        \mathcal R \hat{\mathcal{C}}^{(-)}_{\ell} \mathcal R^{\dagger} &= \mathcal R\left[\textstyle\sum\limits_{j} \krondelwide f_{j-1}^{\dagger}\bar{n}_{j}f_{j+1}\right]\mathcal R^{\dagger}
        \\&= \textstyle\sum\limits_{j'}\delta_{{\textstyle\sum\nolimits_{k=1}^{j'-1}}{\scriptstyle\overline{n}_k,M-\ell-1}} f_{j'+2}^{\dagger}\bar{n}_{j'+1}f_{j'}
        = \hat{\mathcal{C}}^{(+)}_{M-\ell}\,.
    \end{aligned}
    \label{eqn:mapClm}\raisetag{36pt}
\end{align}
Using Eqs.~\eqref{eqn:CLdecop_t}--\eqref{eqn:thetato0}, \eqref{eqn:operator_symmetric}, \eqref{eqn:mapCl0}--\eqref{eqn:mapClm}, we obtain 
\begin{align}
    \begin{split}
        \mathcal C_\ell^{(0)}(t) &= \mathcal C_{M-\ell}^{(0)}(t)\,,
    \\\mathcal C_\ell^{(+)}(\theta, t) &= e^{+i\theta} \mathcal C_{M-\ell}^{(-)}(0,t) =  \mathcal C_{M-\ell}^{(-)}(-\theta,t) \,,
    \\\mathcal C_\ell^{(-)}(\theta, t) &= e^{-i\theta} \mathcal C_{M-\ell}^{(+)}(0,t) =  \mathcal C_{M-\ell}^{(+)}(-\theta,t) \,.
    \end{split}
    \label{eqn:Clmapfin}
\end{align}
At $\theta =0$ this gives $\mathcal C_{M-\ell}^{(\pm )}(0,t)
=
\left[\mathcal C_\ell^{(\pm)}(0,t)\right]^*$. Using this relation, and Eq.~\eqref{eqn:Clmapfin}, we obtain 
\begin{align}
    \mathcal C_{M-\ell}(\theta,t) =  \mathcal C_{M-\ell}^{(0)}(t) + e^{+i\theta}\left[\mathcal C_\ell^{(+)}(0,t)\right]^* + e^{-i\theta}\mathcal C_\ell^{(+)}(0,t)\,.
\end{align}
Since reflection maps spin-chain bond $\ell$ onto bond $M-\ell$, the coefficients $\mathcal C_\ell(\theta,t)$ and $\mathcal C_{M-\ell}(\theta,t)$ will be used to extract the symmetric and antisymmetric parts as
\begin{align}
\begin{split}
    \mathcal C^{(s)}_\ell(\theta,t)&=  \frac{
\mathcal C_\ell(\theta,t)
+
\mathcal C_{M-\ell}(\theta,t)
}{2}
\\&=
\mathcal C_\ell^{(0)}(t)
+
2\Re[\mathcal C_\ell^{(+)}(0,t)]\cos\theta,
\\
\mathcal C^{(a)}_\ell(\theta,t)&=  \frac{
\mathcal C_\ell(\theta,t)
-
\mathcal C_{M-\ell}(\theta,t)
}{2}
\\&=
-2\Im[\mathcal C_\ell^{(+)}(0,t)]\sin\theta.
\end{split}
\end{align}

\noindent The symmetric and antisymmetric decomposition distinguishes the two possible effects of fractional statistics. As we shall see, both the ground-state and prepared initial charge wave functions satisfy $\Im[\mathcal C_\ell^{(+)}(0,0)]=0$ [Figs.\hyperref[fig:Cl_decomposition]{~\ref{fig:Cl_decomposition}} and~\ref{fig:spinexchange_clcoeff}]. The antisymmetric term thus vanishes at $t=0$, confining the statistical dependence to the symmetric $\cos\theta$ contribution. For the ground state, this gives the weak statistical dependence discussed below. For the dynamical initial state, however, subsequent time evolution can generate nonzero $\Im[\mathcal C_\ell^{(+)}(0,t)]$, producing an additional antisymmetric contribution.
 
\begin{figure}[t]
\centering
\includegraphics{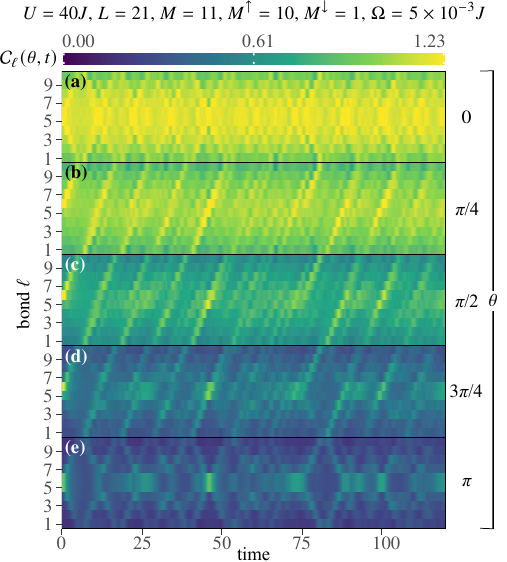}
\caption{\label{fig:dynclcoeff}\textbf{Fractional exchange statistics weakens and skews the exchange pathways governing impurity transport.} Time-dependent GESC coefficients $\mathcal C_\ell(\theta,t)$, following release of a single centrally pinned impurity, governing the local exchange energy scale $J^2\mathcal C_\ell/U$.  $\mathcal C_\ell(\theta,t)$ are shared by the fermion- and boson-type calculations because their charge preparation and evolution are identical. Parameters and release protocol are as in \hyperref[fig:denspinchain]{Fig.~\ref{fig:denspinchain}}. Colors saturate at the 99th percentile. Time is expressed in units of $J^{-1}$.\vspace{36pt}}
\end{figure}

To determine how the antisymmetric contribution emerges from an initially symmetric charge state, we expand the spin-exchange coefficients at short times. The short-time evolution of an operator $\hat O$  up to second order in time is
\begin{widetext}
\vspace{-10pt}
\begin{align}
\expval{\hat O(t)}
=
\expval{\hat O}{\varphi(0)}
+
it\expval{[H_{\mathrm C},\hat O]}{\varphi(0)}
-
\frac{t^2}{2}
\expval{[H_{\mathrm C},[H_{\mathrm C},\hat O]]}{\varphi(0)}
+
\order{t^3},
\label{eq:HeisenbergShort}
\end{align}
Applying Eq.~\eqref{eq:HeisenbergShort} to Eq.~\eqref{eqn:CLdecop_t} at $\theta=0$ gives
\begin{align}
    \begin{split}
        \mathcal C_\ell^{(+)}(0,t) &= \mathcal C_\ell^{(+)}(0,0)
        + it \expval{ \comm{H_{\mathrm C}}{\hat{\mathcal C}_\ell^{(+)}} }{\varphi(0)}- \frac{t^2}{2}
        \expval{
            \comm{H_{\mathrm C}}{ \comm{H_{\mathrm C}}{\hat{\mathcal C}_\ell^{(+)}}}
        }{\varphi(0)} + \order{t^3}\,, \\
        \mathcal C_\ell^{(-)}(0,t) = [\mathcal C_\ell^{(+)}(0,t)]^* &= \mathcal C_\ell^{(+)}(0,0)
        - it \expval{ \comm{H_{\mathrm C}}{\hat{\mathcal C}_\ell^{(+)}} }{\varphi(0)}- \frac{t^2}{2}
        \expval{
            \comm{H_{\mathrm C}}{ \comm{H_{\mathrm C}}{\hat{\mathcal C}_\ell^{(+)}}}
        }{\varphi(0)} + \order{t^3}\,, \\
        \mathcal C_\ell^{(0)}(t) &= \mathcal C_\ell^{(0)}(0) - \frac{t^2}{2} \expval{
            \comm{H_{\mathrm C}}{ \comm{H_{\mathrm C}}{\hat{\mathcal C}_\ell^{(0)}}}
        }{\varphi(0)} + \order{t^3}\,.
    \end{split}
    \label{eqn:cl0t}
\end{align}
$\mathcal C_\ell^{(0)}(0,t)$ has no linear-in-time term for the charge Hamiltonian considered here. Eq.~\eqref{eqn:cl0t} show that although $\mathcal C_\ell^{(\pm)}(0,0)$ are purely real, $\mathcal C_\ell^{(\pm)}(0,t)$ acquires an imaginary part linear in time. The resulting antisymmetric contribution remains zero at $\theta=0$ and $\pi$ on account of $\sin\theta$ being $0$, but becomes finite at intermediate $\theta$, consistent with \hyperref[fig:dynclcoeff]{Fig.~\ref{fig:dynclcoeff}}. Using Eqs.~\eqref{eqn:thetato0} and \eqref{eqn:cl0t},
\begin{align}
    \begin{split}
        \mathcal C_\ell(\theta,t) = \mathcal C_\ell^{(0)}(0) &- \frac{t^2}{2} \expval{
            \comm{H_{\mathrm C}}{ \comm{H_{\mathrm C}}{\hat{\mathcal C}_\ell^{(0)}} }
        }{\varphi(0)} \\&+ e^{+i\theta} \left[
            \begin{aligned}
                \mathcal C_\ell^{(+)}(0,0) &+ it \expval{ \comm{H_{\mathrm C}}{\hat{\mathcal C}_\ell^{(+)}} }{\varphi(0)}
                - \frac{t^2}{2} \expval{
                    \comm{H_{\mathrm C}}{ \comm{H_{\mathrm C}}{\hat{\mathcal C}_\ell^{(+)}} }
                }{\varphi(0)}
            \end{aligned}
        \right] \\
        &+ e^{-i\theta} \left[
            \begin{aligned}
                &\mathcal C_\ell^{(+)}(0,0) - it \expval{ \comm{H_{\mathrm C}}{\hat{\mathcal C}_\ell^{(+)}} }{\varphi(0)}
                - \frac{t^2}{2} \expval{
                    \comm{H_{\mathrm C}}{ \comm{H_{\mathrm C}}{\hat{\mathcal C}_\ell^{(+)}} }
                }{\varphi(0)}
            \end{aligned}
        \right] + \order{t^3}.
    \end{split}
    \label{eq:Ctheta_before_simplify}
\end{align}
Simplifying and retaining terms till the second order:
\begin{align}
\mathcal C_\ell(\theta,t)
=
\mathcal C_\ell^{(0)}(0)+2\mathcal C_\ell^{(+)} (0,0)\cos\theta \notag
&-
\frac{t^2}{2}
\expval{
\comm{H_{\mathrm C}}{
\comm{H_{\mathrm C}}{\hat{\mathcal C}_\ell^{(0)}}}
}{\varphi(0)}\notag
\\&-
t^2\expval{
\comm{H_{\mathrm C}}{
\comm{H_{\mathrm C}}{\hat{\mathcal C}_\ell^{(+)}}}
}{\varphi(0)}\cos\theta
-2t\expval{
\comm{H_{\mathrm C}}{\hat{\mathcal C}_\ell^{(+)}}
}{\varphi(0)}\sin\theta
\,.
\label{eq:Ctheta_simplify}
\end{align}
\end{widetext}
For the ground state, setting $t=0$ gives
\begin{align}
\mathcal C_\ell(\theta)
=
\mathcal C_\ell^{(0)}
+
2\mathcal C_\ell^{(+)}\cos\theta.
\label{eq:Ctheta_initial}
\end{align}
Thus, the ground-state
profile is reflection symmetric,
\begin{align}
\mathcal C_\ell(\theta)
=
\mathcal C_{M-\ell}(\theta)\,.
\label{eq:Ctheta_initial1}
\end{align}
As $\theta$ goes from $0$ to $\pi$,
\begin{align}
    \mathcal C_\ell^{(0)}
+
2\mathcal C_\ell^{(+)} \rightarrow \mathcal C_\ell^{(0)}
-
2\mathcal C_\ell^{(+)}\,.
\label{eq:Ctheta_initial2}
\end{align}

From Eqs.~\eqref{eq:Ctheta_initial}, \eqref{eq:Ctheta_initial1}, and \eqref{eq:Ctheta_initial2}, the ground-state spin-exchange profile has no antisymmetric contribution; fractional statistics only rescales the coefficients through the symmetric $\cos\theta$ term. This is consistent with \hyperref[fig:Cl_decomposition]{Fig.~\ref{fig:Cl_decomposition}}, where $\mathcal C_\ell(\theta)$ retains essentially the same bond dependence as $\theta$ varies while its overall magnitude changes. The statistical phase therefore acts predominantly as a renormalization of the exchange scale, which explains why the spin ground state is only weakly reconstructed. \hyperref[fig:spinexchange_clcoeff]{Figure~\ref{fig:spinexchange_clcoeff}} shows that both the magnitude of $\mathcal{C}_{\ell}(\theta)$ and its variation with $\theta$ increase with filling $M/L$. At fixed $M$ and $L$, $\mathcal{C}_{\ell}(\theta)$ has no explicit $N$ dependence because it is determined entirely by the charge wave function. Along the ground-state sequence with fixed $M/N=6$, however, increasing $N$ also increases $M/L$. The stronger exchange scale and statistical modulation therefore arise indirectly through the filling $M/L$ rather than directly from the number of components $N$. Thus, although the statistical modulation becomes larger, the spin ground state remains only weakly reconstructed.

The dynamics remains more sensitive to $\theta$ than the ground state. As discussed above, even for a reflection-symmetric initial charge state, an antisymmetric component of $\mathcal C_\ell(\theta,t)$ can develop during the evolution, while the symmetric contribution remains present. Thus, at finite time, both contributions enter the spin-exchange coefficients
\begin{align}
\begin{split}
    \mathcal C^{(s)}_\ell(\theta,t)= &\;
 \mathcal C_\ell^{(0)}(0)
+
2\mathcal C_\ell^{(+)}(0,0)\cos\theta
\\& -
\frac{t^2}{2}
\expval{
\comm{H_{\mathrm C}}{
\comm{H_{\mathrm C}}{\hat{\mathcal C}_\ell^{(0)}}}
}{\varphi(0)}
\\&-t^2
\expval{
\comm{H_{\mathrm C}}{
\comm{H_{\mathrm C}}{\hat{\mathcal C}_\ell^{(+)}}}
}{\varphi(0)}\cos\theta,
\\
\mathcal C^{(a)}_\ell(\theta,t)
=&
-2t
\expval{
\comm{H_{\mathrm C}}{\hat{\mathcal C}_\ell^{(+)}}
}{\varphi(0)}\sin\theta.\raisetag{16pt}
\end{split}
\end{align}
These equations provide the analytical distinction between the static and dynamical cases. The symmetric $\cos\theta$ contribution survives at finite time and continues to reduce the overall exchange scale as $\theta$ increases, while the dynamically generated imaginary part of $\mathcal C_\ell^{(+)}(0,t)$ produces the additional $\sin\theta$ antisymmetric contribution. The short-time expansion makes the origin of this asymmetry explicit: it is generated by the commutator of the directed three-site term with the charge Hamiltonian and appears at linear order in time.

\hyperref[fig:dynclcoeff]{Figure~\ref{fig:dynclcoeff}} shows both effects directly following release of the centrally pinned impurity. The exchange profile propagates across the spin chain and subsequently refocuses. Its overall magnitude decreases with increasing $\theta$, consistent with the surviving symmetric contribution, while intermediate statistical phases develop a pronounced right--left asymmetry, as predicted by the antisymmetric $\sin\theta$ term. The asymmetry disappears at the statistical endpoints, where $\sin\theta=0$. For the common charge initial state and evolution used here, the coefficients $\mathcal C_\ell(\theta,t)$ are identical for fermion- and boson-type anyons; the distinction between the two enters through the spin operators multiplying these coefficients in their respective GESC Hamiltonians.

The consequence of this dynamical exchange profile for transport properties can be made explicit using the fermion-type spin-chain Hamiltonian, Eq.~\eqref{eqn:HSCF}, as an example. The Heisenberg equation for the impurity occupation
is
\begin{align}
\dot{n}_{\ell,\downarrow}
=
\mathcal J_{\ell-1,\ell}
-
\mathcal J_{\ell,\ell+1}\,,
\label{eqn:impurity_continuity}
\end{align}
where the impurity exchange current across bond $(\ell,\ell+1)$ is $\mathcal J_{\ell,\ell+1}
=
iJ^2\mathcal C_\ell(\theta,t)
\left(
\sigma_\ell^-\sigma_{\ell+1}^+
-
\sigma_\ell^+\sigma_{\ell+1}^-
\right)/U$.
Equation~\eqref{eqn:impurity_continuity} shows that $\mathcal C_\ell(\theta,t)$ directly controls the impurity exchange current across bond $\ell$. To connect this to the evolution of the impurity along the spin chain, we consider an impurity localized at spin-chain site $\ell_0$ at time $t_0$. Over a short interval $\delta t$, the first nonvanishing transfer to the neighboring spin-chain sites is quadratic in $\delta t$,
{\medmuskip=0mu
 \thinmuskip=0mu
\begin{align}
\begin{split}
    &\expval{n_{\ell_0,\downarrow}(t_0+\delta t)}
=
1-
\frac{J^4}{U^2}\left[
\mathcal C_{\ell_0-1}^2(\theta,t_0)
+
\mathcal C_{\ell_0}^2(\theta,t_0)
\right]\delta t^2
+\order{\delta t^3}\,,
\\&\expval{n_{\ell_0-1,\downarrow}(t_0+\delta t)}
=
\frac{J^4}{U^2} \mathcal C_{\ell_0-1}^2(\theta,t_0)\delta t^2
+\order{\delta t^3},
\\&\expval{n_{\ell_0+1,\downarrow}(t_0+\delta t)}
=
\frac{J^4}{U^2} \mathcal C_{\ell_0}^2(\theta,t_0)\delta t^2
+\order{\delta t^3}.
\end{split}
\label{eqn:impurity_transfer}
\end{align}
}

\begin{figure*}
\centering
\includegraphics{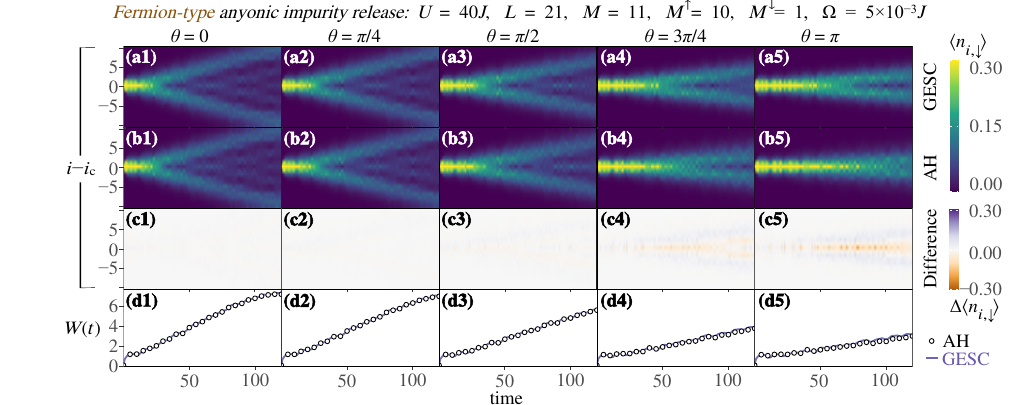}
\caption{\label{fig:gesc_benchmark_IR}\textbf{GESC captures the statistical suppression of impurity propagation.} Rowwise: the fermion-type impurity density $\langle n_i^\downarrow(t)\rangle$ from GESC \hyperref[fig:denspinchain]{[\textit{id.}~Fig.~\ref{fig:denspinchain}(a)]}, and that from the full Anyon--Hubbard (AH) model, their signed difference (GESC minus AH), and the RMS displacement $W(t)$. The central pinning field $h_z=8J$ is removed at $t=0$, while the weak harmonic confinement remains. Viridis colors saturate at the 99th percentile. Time is expressed in units of $J^{-1}$.}
\end{figure*}

From Eq.~\eqref{eqn:impurity_transfer}, the local right--left imbalance is therefore
\begin{align}
\begin{split}
&\expval{n_{\ell_0+1,\downarrow}(t_0+\delta t)}
-
\expval{n_{\ell_0-1,\downarrow}(t_0+\delta t)}
\\
&=
\left(\frac{J^2}{U}\right)^2
\left[
\mathcal C_{\ell_0}^2(\theta,t_0)
-
\mathcal C_{\ell_0-1}^2(\theta,t_0)
\right]
\delta t^2
+\order{\delta t^3}.
\label{eqn:impurity_short_asymmetry}
\end{split}
\end{align}
Equations~\eqref{eqn:impurity_transfer} and \eqref{eqn:impurity_short_asymmetry}, respectively, show that the neighboring $\mathcal C_\ell$ coefficients determine the short-time transfer away from $\ell_0$ and its right--left difference. This nearest-neighbor result corresponds to the $r=1$ exchange process. To extend this result beyond nearest neighbors, consider the transfer to a site $r$ positions away. Reaching $\ell_0\pm r$ requires at least $r$ successive exchanges, $\ell_0\rightarrow\ell_0\pm 1\rightarrow\ldots\rightarrow\ell_0\pm r$. The leading contribution to the corresponding amplitude is
\begin{widetext}
\begin{align}
\begin{split}
a_{\ell_0+r}(t)
=
(-i)^r&\left(\frac{J^2}{U}\right)^r
\int\limits_{\mathrlap{\!\!t_0<t_1<\ldots<t_r<t}}\big[
dt_1\ldots dt_r \;
\mathcal C_{\ell_0+r-1}(\theta,t_r)
\ldots
\mathcal C_{\ell_0+1}(\theta,t_2)
\mathcal C_{\ell_0}(\theta,t_1)\big]\,,
\\a_{\ell_0-r}(t)
=
(-i)^r&\left(\frac{J^2}{U}\right)^r
\int\limits_{\mathrlap{\!\!t_0<t_1<\ldots<t_r<t}}\big[
dt_1\ldots dt_r \;
\mathcal C_{\ell_0-r}(\theta,t_r)
\ldots
\mathcal C_{\ell_0-2}(\theta,t_2)
\mathcal C_{\ell_0-1}(\theta,t_1)\big]\,.
\end{split}
\label{eq:impurity_multistep}
\end{align}
\end{widetext}
where
$\expval{n_{\ell_0\pm r,\downarrow}(t)}
\simeq
\abs{a_{\ell_0 \pm r}(t)}^2$.
Equations~\eqref{eq:impurity_multistep} show that propagation over an increasing distance involves products of an increasing number of $\mathcal C_\ell(\theta,t)$ coefficients. The overall reduction of $\mathcal C_\ell(\theta,t)$ with increasing $\theta$ therefore suppresses these multistep contributions and hence suppresses long-distance propagation. The right- and left-moving amplitudes in Eq.~\eqref{eq:impurity_multistep} sample different sequences of $\mathcal C_\ell(\theta,t)$. Consequently, the nearest-neighbor asymmetry identified in Eq.~\eqref{eqn:impurity_short_asymmetry} can accumulate over successive exchanges: at intermediate $\theta$, the spatially asymmetric $\mathcal C_\ell(\theta,t)$ profile produces increasingly different right- and left-moving amplitudes, and hence a pronounced difference between $\langle{n_{\ell_0+r,\downarrow}}\rangle$ and $\langle{n_{\ell_0-r,\downarrow}}\rangle$. This provides a direct microscopic connection between the spatiotemporal spin-exchange coefficients in \hyperref[fig:dynclcoeff]{Fig.~\ref{fig:dynclcoeff}} and the slower, inversion-asymmetric impurity transport. Dynamical enhancement of state-mediated statistical dependence therefore does not require an explicit anyonic string in the measured impurity occupation; it follows from the repeated action of the statistics-dependent spin-exchange Hamiltonian during the evolution.

\begin{figure*}
\centering
\includegraphics{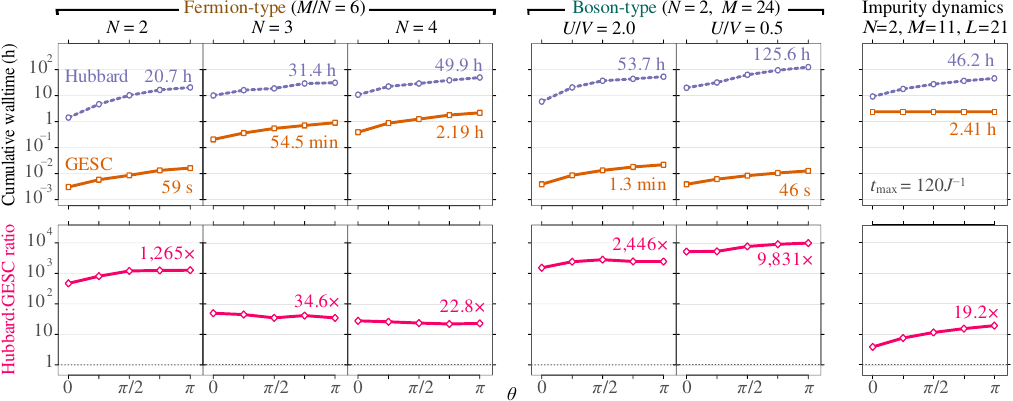}
\caption{\label{fig:gesc_walltime_unscaled}\textbf{Spin--charge separation reduces the cost of strong-interaction computations and enables resolved statistical-phase scans.} Cumulative wall-clock times (upper row) and Hubbard:GESC time ratios (lower row) for ground-state correlations and (last column)
impurity dynamics. Costs accumulate through the indicated statistical phase $\theta$. Correlation jobs were allocated 20 cores each; dynamics jobs used 24 cores for Hubbard and 20 and 4 for the GESC charge and spin calculations, respectively. 
Sensitivity analysis accounting for processor heterogeneity results in a negligible clock-only rescaling of the cumulative ratios by $-3.2\%$ to $+4.2\%$ and leaves the dynamics ratio unchanged.}
\end{figure*}

\section{\label{app:gesc_dynamics_benchmarks}Benchmarking the GESC dynamics}

\hyperref[fig:gesc_benchmark_IR]{Figure~\ref{fig:gesc_benchmark_IR}} benchmarks the fermion-type GESC impurity dynamics against the full Anyon--Hubbard (AH) evolution. The GESC reproduces the overall suppression of impurity propagation as $\theta$ increases, including the propagating density fronts and the enhanced central localization at larger statistical phase. The agreement is clear in the RMS displacement $W(t)$, for which the GESC and AH results overlap for all $\theta$. The residual density differences are small and become most visible near the trap center at late times for the larger statistical phases. At $\theta=\pi$, these deviations are largest, with the GESC slightly overestimating the impurity width. Overall, the benchmark demonstrates that the GESC captures both the spatially resolved impurity propagation and its statistics-dependent suppression in the strongly interacting regime.

\section{\label{app:gesc_efficiency}Computational efficiency}
The GESC formalism reduces computational cost by separating the strongly interacting problem into a spinless charge problem and an effective spin chain. This avoids the coupled charge--spin calculation required by the full Hubbard model. The resulting computational advantage depends on the system size, number of components, convergence requirements, and observables being evaluated. The full GESC workflow comprises the charge calculation, construction of the bond-dependent spin-exchange coefficients, solution of the spin problem, and evaluation of the selected observables. The Hubbard workflow comprises ground-state preparation and evaluation of the same observables. For the dynamical benchmark, both workflows also include time evolution.

GESC shows substantial computational advantage for both ground-state properties and impurity dynamics, with the largest gains in the boson-type benchmarks. Statistical-phase scans especially benefit from reusing the $\theta$-independent charge calculation. This advantage is clearest for impurity dynamics: the shared charge evolution dominates the GESC cost, while each additional phase requires only $2$~s, compared with approximately $8.5$ to $9$~h for Hubbard. Including the shared charge calculation, GESC reduces computing time by about $95\%$ across the five statistical phases shown here, with greater relative savings as more phases are sampled. The nearly flat cumulative GESC time shows that much finer phase sampling is accessible at negligible additional cost.

Calculations used Intel\textregistered~Xeon\textregistered~processors spanning Haswell, Cascade Lake, Ice Lake, and Sapphire Rapids architectures, together with AMD EPYC Rome processors. Accounting for the hardware heterogeneity, the stated clock-only rescaling changes the cumulative ratios shown in the bottom row of \hyperref[fig:gesc_walltime_unscaled]{Fig.~\ref{fig:gesc_walltime_unscaled}} negligibly as detailed in the figure caption.
\bibliography{references}
\end{document}